\documentclass[%
 reprint,
superscriptaddress,
nofootinbib,
 amsmath,amssymb,
 aps,
]{revtex4-2}

\usepackage{graphicx}
\usepackage{dcolumn}
\usepackage{bm}

\usepackage[justification=justified]{caption}
\usepackage{subcaption}
\usepackage{ragged2e}

\DeclareCaptionJustification{justified}{\justifying}

\begin{document}

\preprint{APS/123-QED}

\title{A Platform for Braiding Majorana Modes with Magnetic Skyrmions}

\author{Shiva T. Konakanchi}
 \email{skonakan@purdue.edu}
\affiliation{Department of Physics and Astronomy, Purdue University, West Lafayette, Indiana 47907, USA}

\author{Jukka I. V\"ayrynen}
\affiliation{Department of Physics and Astronomy, Purdue University, West Lafayette, Indiana 47907, USA}

\author{Yong P. Chen}
\affiliation{Department of Physics and Astronomy, Purdue University, West Lafayette, Indiana 47907, USA}
\affiliation{Elmore Family School of Electrical and
Computer Engineering, Purdue University, West Lafayette,
Indiana 47907, USA}
\affiliation{Birck Nanotechnology Center, and Purdue
Quantum Science and Engineering Institute, Purdue
University, West Lafayette, Indiana 47907, USA}

\author{Pramey Upadhyaya}
\affiliation{Elmore Family School of Electrical and
Computer Engineering, Purdue University, West Lafayette,
Indiana 47907, USA}

\author{Leonid P. Rokhinson}
\email{leonid@purdue.edu}
\affiliation{Department of Physics and Astronomy, Purdue University, West Lafayette, Indiana 47907, USA}
\affiliation{Elmore Family School of Electrical and
Computer Engineering, Purdue University, West Lafayette,
Indiana 47907, USA}
\affiliation{Birck Nanotechnology Center, and Purdue
Quantum Science and Engineering Institute, Purdue
University, West Lafayette, Indiana 47907, USA}

\date{\today}

\begin{abstract}
After a decade of intense theoretical and experimental efforts, demonstrating braiding of Majorana modes remains an unsolved problem in condensed matter physics due to platform specific challenges. In this work, we propose topological superconductor – magnetic multilayer heterostructures with on-chip microwave cavity readout as a novel platform for initializing, braiding and reading out Majorana modes. Stray fields from a skyrmion in the magnetic layers can nucleate a vortex in the topological superconducting layer. Such a vortex is known to host Majorana bound states at its core. Through analytical calculations within London and Thiele formalisms, and through micromagnetic simulations, we show that our nucleation and braiding scheme can be effectively realized with a variety of existing options for magnetic and superconducting layers. Further, we show that the coupling of the Majorana bound states to electric field of a resonator leads to an experimentally observable parity-dependent dispersive shift of the resonator frequency. Our work paves the way for realizing Majorana braiding in the near future.
\end{abstract}

\maketitle


\section{\label{sec:intro}Introduction}
 
Demonstration of non-abelian exchange statistics is one of the most active areas of condensed matter research and yet experimental realization of braiding of Majorana modes remains elusive~\cite{RevModPhys.80.1083,zhang2019next}. Most efforts so far have been focused on superconductor/semiconductor nanowire hybrids, where Majorana bound states (MBS) are expected to form at the ends of a wire or at boundaries between topologically trivial and non-trivial regions~\cite{rokhinson2012fractional, deng2012anomalous, mourik2012signatures, LutchynReview}. Recently, it became clear that abrupt interfaces may also host topologically trivial Andreev states with experimental signatures similar to MBS \cite{pan2020generic,Yu2021}, which makes demonstrating braiding in nanowire-based platforms challenging. Phase-controlled long Josephson junctions (JJ) open much wider phase space to realize MBS with a promise to solve some problems of the nanowire platform, such as enabling zero-field operation to avoid detrimental flux focusing for in-plane fields \cite{pientka2017topological, ren2019topological}. However, MBSs in long JJs suffer from the same problems as in the original Fu-Kane proposal for topological insulator/superconductor JJs, such as poor control of flux motion along the junction and presence of sharp interfaces in the vicinity of MBS-carrying vortices which may host Andreev states and trap quasiparticles. For instance, MBS spectroscopy in both HgTe and InAs-based JJs shows a soft gap \cite{fornieri2019evidence}, despite a hard SC gap in an underlying InAs/Al heterostructure.

\begin{figure*}[t]
\centering
\begin{subfigure}{0.95\textwidth}
\includegraphics[width=1\textwidth]{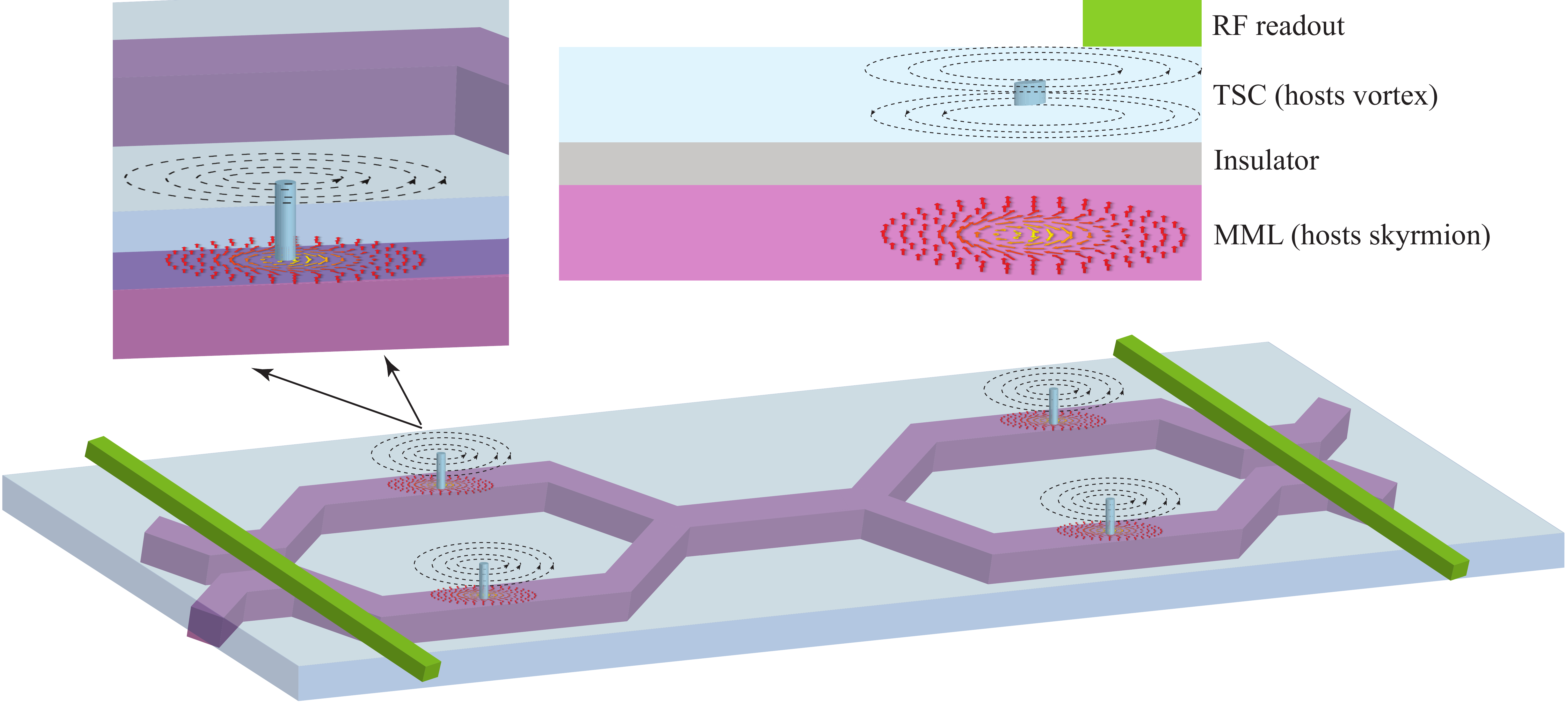}
\caption{\label{fig:schematic}}
\end{subfigure}
\begin{subfigure}{0.35\textwidth}
\includegraphics[width=1\textwidth]{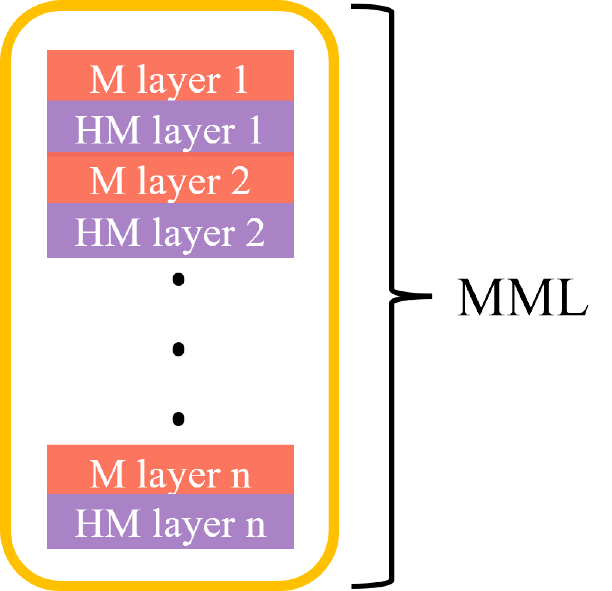}
\caption{\label{fig:layers}}
\end{subfigure}
\begin{subfigure}{0.6\textwidth}
\includegraphics[width=1\textwidth]{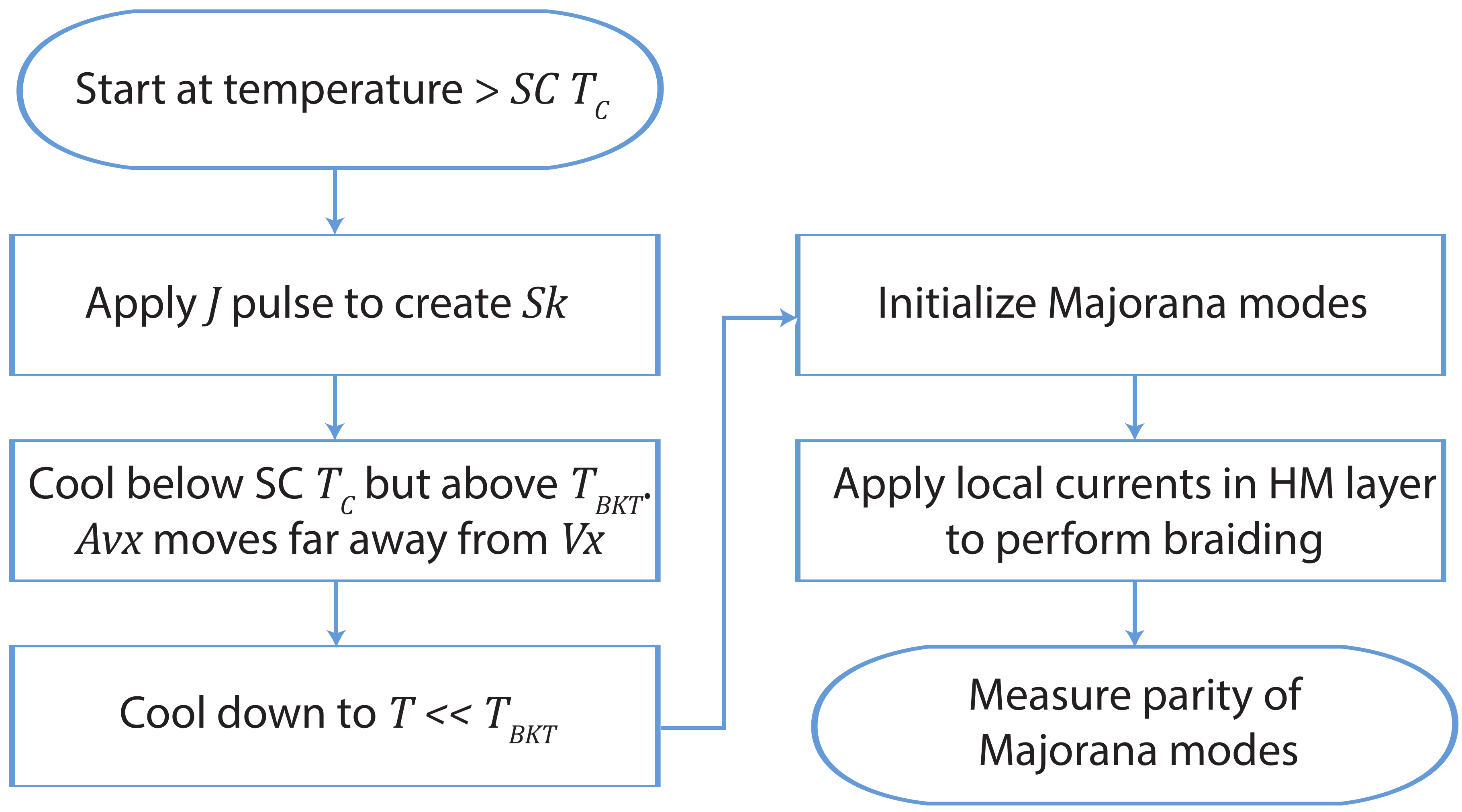}
\caption{\label{fig:flow}}
\end{subfigure}
\caption{\label{fig:one} (a) Schematic of the Majorana braiding platform. Magnetic multilayer (MML) is patterned into a track and is separated from TSC by a thin insulating layer. Green lines represent on-chip microwave resonators for a dispersive parity readout setup. The left inset shows a magnified view of a SVP and the right inset shows the role of each layer (b) Expanded view of the composition of an MML (c) Process flow diagram for our Majorana braiding scheme. Here, $T_c$ is superconducting transition temperature and $T_{BKT}$ is Berezinskii–Kosterlitz–Thouless transition temperature for the TSC.}

\end{figure*}

In the search for alternate platforms to realize Majorana braiding, spectroscopic signatures of MBS have been recently reported in STM studies of vortex cores in iron-based topological superconductors (TSC) \cite{wang2018evidence}. Notably, a hard gap surrounding the zero-bias peak at a relatively high temperature of $0.55$ K, and a $5$ K separation gap from trivial Caroli-de Gennes-Matricon (CdGM) states were observed \cite{chen2020observation, chen2018discrete}. Moreover, vortices in a TSC can be field-coupled to a skyrmion in an electrically-separated magnetic multilayer (MML) \cite{volkov,petrovic2021skyrmion}, which can be used to manipulate the vortex. This allows for physical separation of the manipulation layer from the layer wherein MBS reside, eliminating the problem of abrupt interfaces faced by nanowire hybrids and JJs. Finally, recent advances in the field of spintronics provide a flexible toolbox to design MML in which skyrmions of various sizes can be stabilized in zero external magnetic field and at low temperatures \cite{petrovic2021skyrmion, buttner2018theory, dupe2016engineering}. Under the right conditions, stray fields from these skyrmions alone can nucleate vortices in the adjacent superconducting layer. In this paper, we propose TSC--MML heterostructures hosting skyrmion-vortex pairs (SVP) as a viable platform to realize Majorana braiding. By patterning the MML into a track and by driving skyrmions in the MML with local spin-orbit torques (SOT), we show that the SVPs can be effectively moved along the track, thereby facilitating braiding of MBS bound to vortices.

The notion of coupling skyrmions (Sk) and superconducting vortices (Vx) through magnetic fields has been studied before \cite{volkov, baumard2019generation, zhou_fusion_2022, PhysRevLett.117.077002, PhysRevB.105.224509, PhysRevB.100.064504, PhysRevB.93.224505, PhysRevB.99.134505, PhysRevApplied.12.034048}. Menezes et al. \cite{menezes2019manipulation} performed numerical simulations to study the motion of a skyrmion--vortex pair when the vortex is dragged via supercurrents and Hals et al. \cite{hals2016composite} proposed an analytical model for the motion of such a pair where a skyrmion and a vortex are coupled via exchange fields. However, the dynamics of a SVP in the context of Majorana braiding remains largely unexplored. Furthermore, no \textit{in-situ} non-demolition experimental technique has been proposed to measure MBS in these TSC--MML heterostructures. In this paper, through micromagnetic simulations and analytical calculations within London and Thiele formalisms, we study the dynamics of a SVP subjected to external spin torques. We demonstrate that the SVP moves without dissociation up to speeds necessary to complete Majorana braiding within estimated quasiparticle poisoning time. We further eliminate the problem of \textit{in-situ} MBS measurements by proposing a novel on-chip microwave readout technique. By coupling the electric field of the microwave cavity to dipole-moments of transitions from Majorana modes to CdGM modes, we show that a topological non-demolition dispersive readout of the MBS parity can be realized. Moreover, we show that our platform can be used to make the first experimental observations of quasiparticle poisoning times in topological superconducting vortices.

The paper is organized as follows: in Section~\ref{sec:plat} we present a schematic and describe our platform. In Section~\ref{sec:initial} we present the conditions for initializing a skyrmion--vortex pair and discuss its equilibrium properties. In particular, we characterize the skyrmion--vortex binding strength. In Section~\ref{sec:braid} we discuss the dynamics of a SVP in the context of braiding. Then in Section~\ref{sec:read}, we present details of our microwave readout technique. Finally, we discuss the scope of our platform in Section~\ref{sec:summ}.

\begin{figure*}[t]
\centering
  \begin{subfigure}{0.32\textwidth}
 \includegraphics[width=1\textwidth]{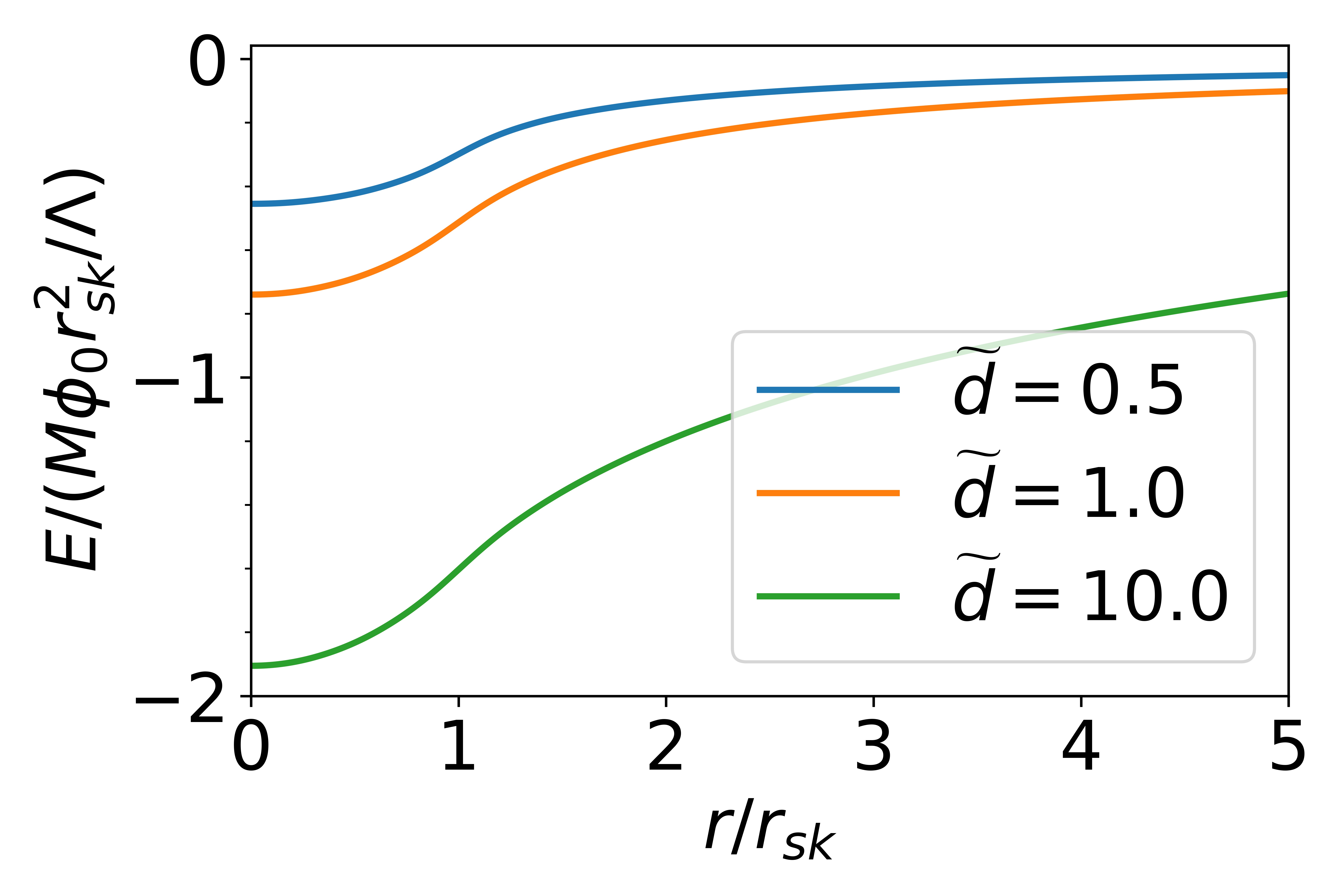}
 \caption{\label{fig:energies}}
 \end{subfigure}
 \begin{subfigure}{0.32\textwidth}
 \includegraphics[width=1\textwidth]{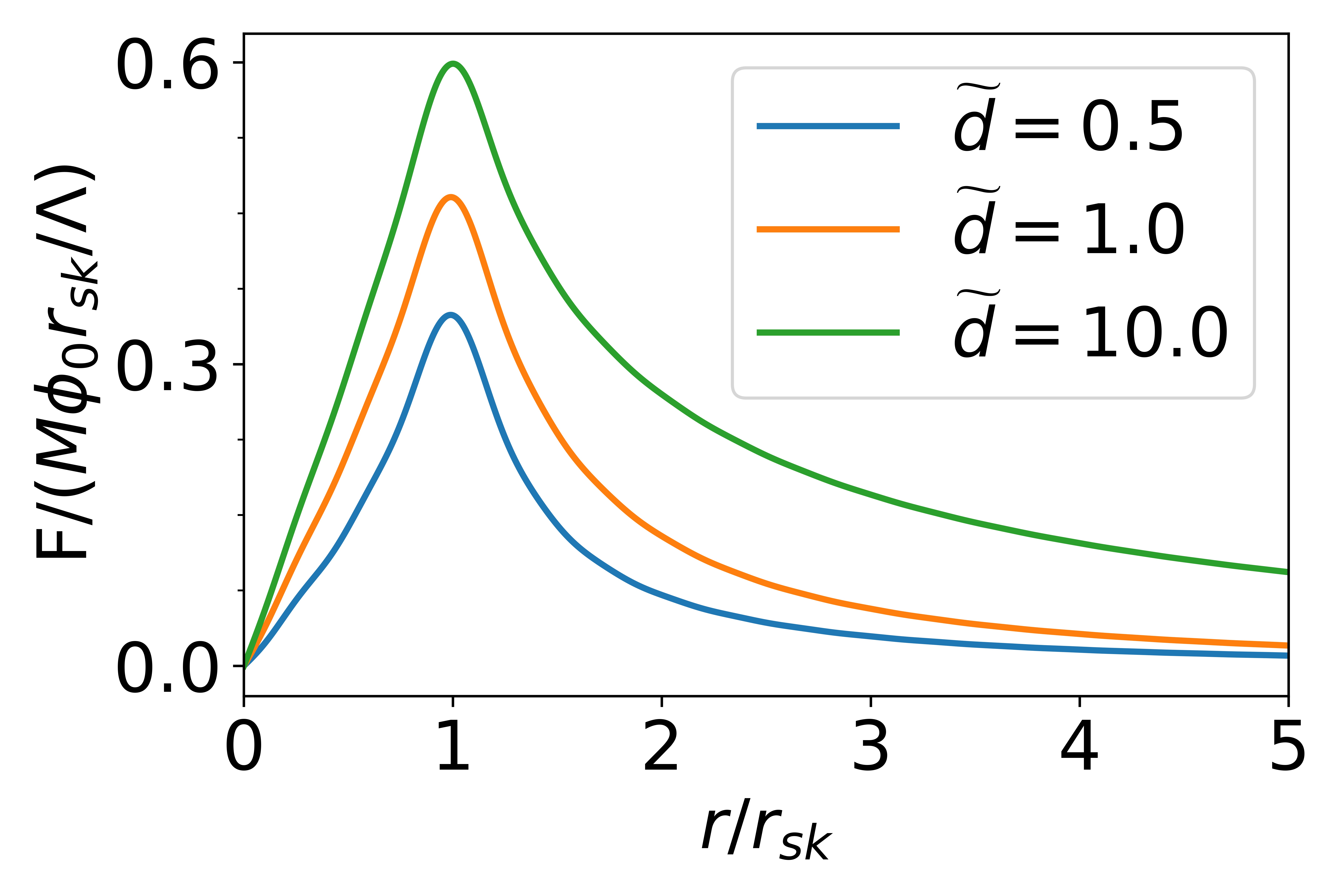}
 \caption{\label{fig:forces}}
 \end{subfigure}
 \begin{subfigure}{0.32\textwidth}
 \includegraphics[width=1\textwidth]{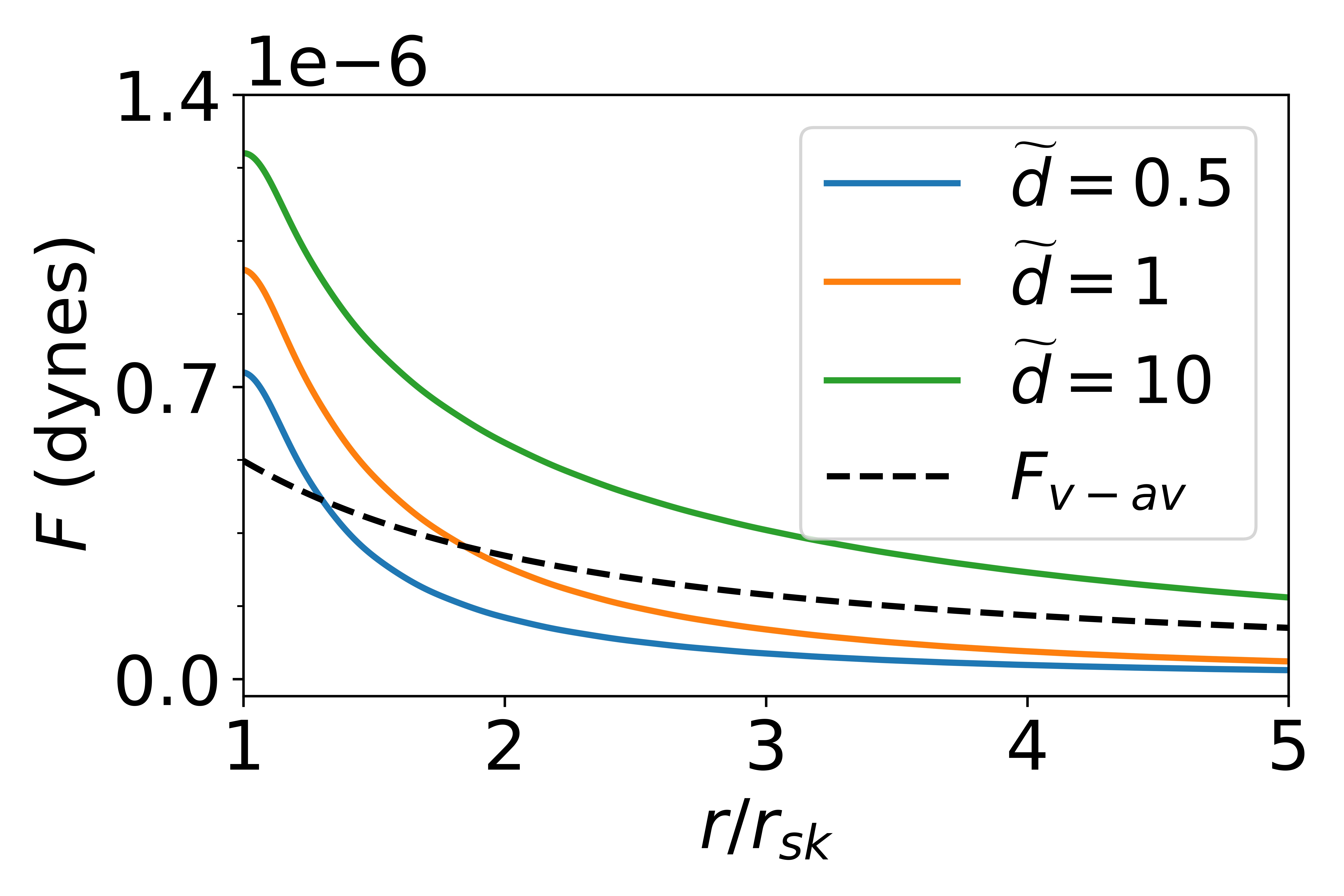}
 \caption{\label{fig:fvav}}
 \end{subfigure}
 \caption{\label{fig:onenew} (a -- b) Normalized energies and forces for Sk--Vx interaction between a Pearl vortex and a N\'eel skyrmion of varying thickness. (c) Attractive $F_{Vx-Avx}$ and repulsive $F_{Sk-Avx}$ (colored lines) for the example materials in Appendix~\ref{app:A}: $M_{0}=1450$ emu/cc, $r_{sk}=35$ nm, $d_s = 50$ nm, $\Lambda = 5$ $\mu$m and $\xi=15$ nm.}

\end{figure*}

\section{\label{sec:plat}Platform Description}

\begin{figure*}[t]
\centering
 \begin{subfigure}{0.59\textwidth}
 \includegraphics[width=1\textwidth]{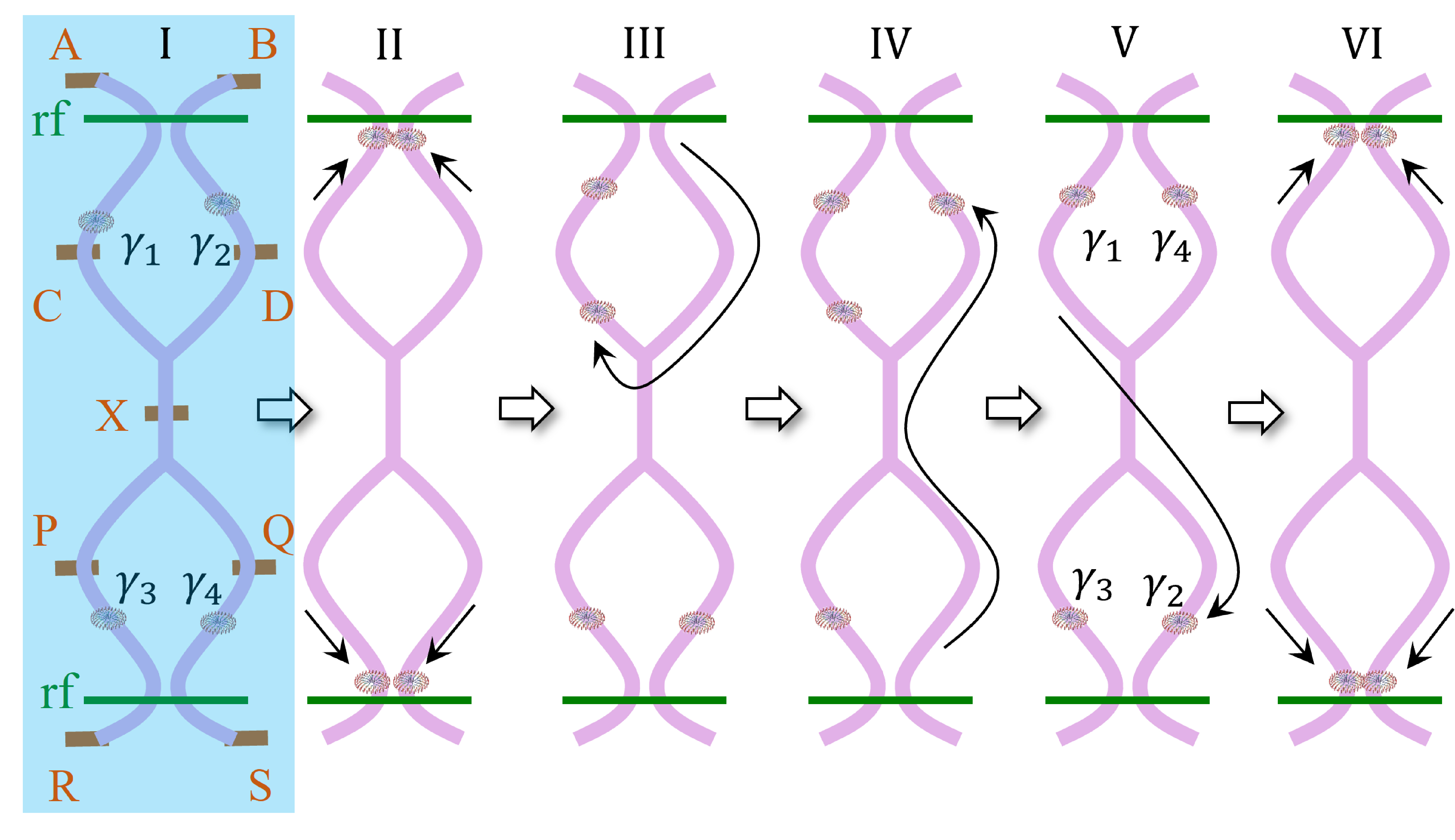}
 \caption{\label{fig:braiding}}
 \end{subfigure}
 \begin{subfigure}{0.39\textwidth}
 \includegraphics[width=1\textwidth]{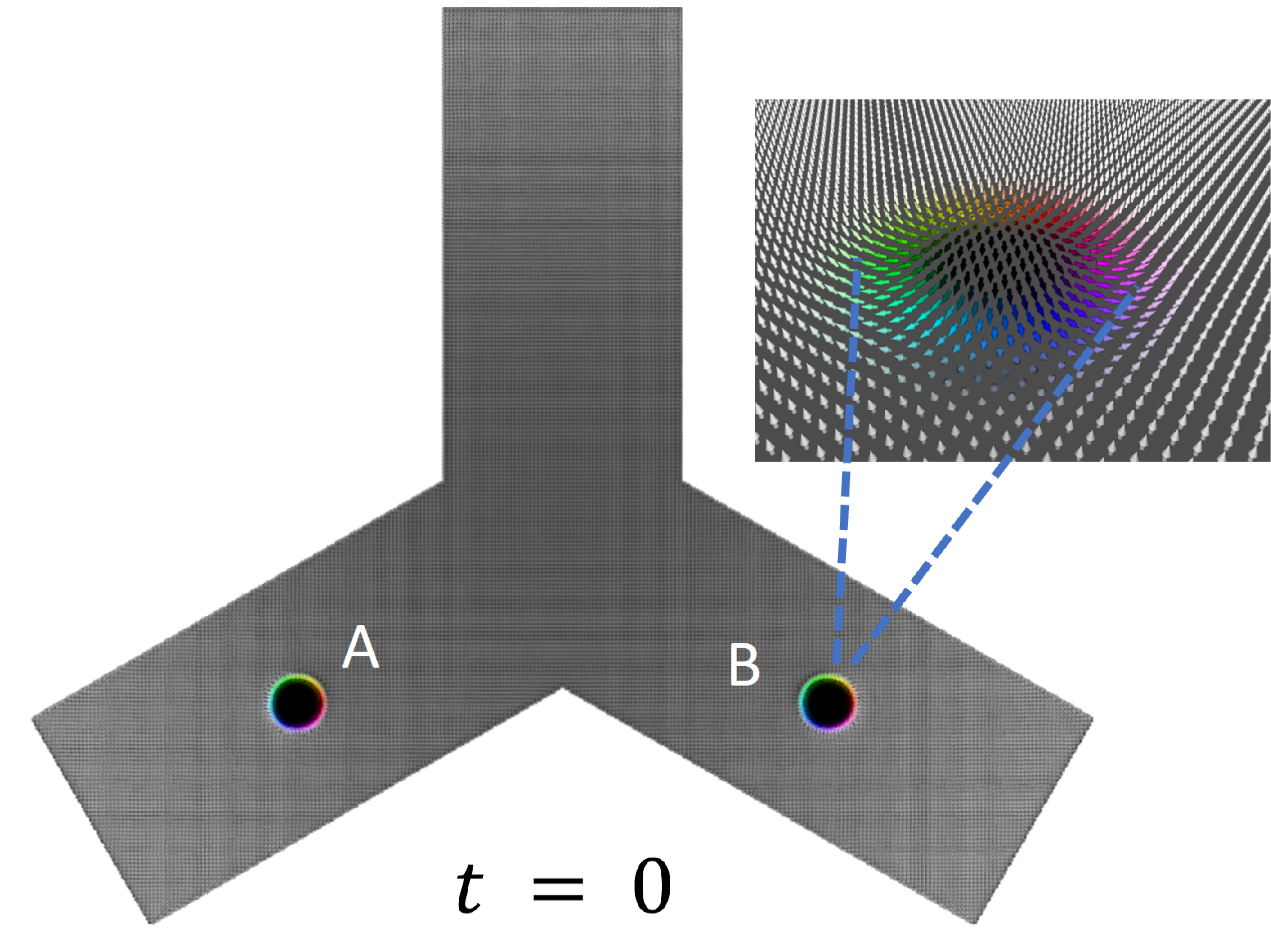}
 \caption{\label{fig:t0}}
 \end{subfigure}
 
 \begin{subfigure}{0.15\textwidth}
 \includegraphics[width=1\textwidth]{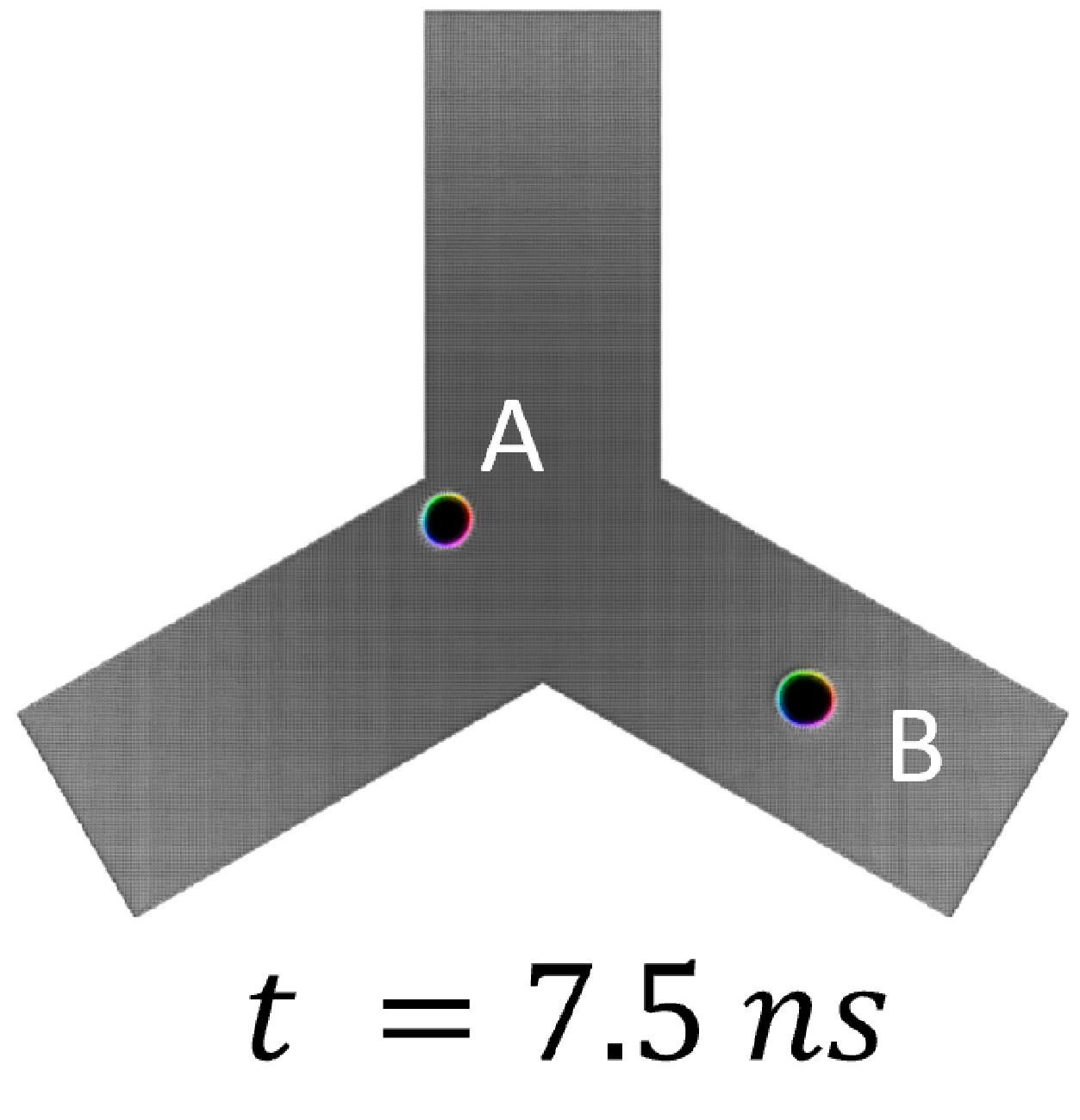}
 \caption{\label{fig:t1}}
 \end{subfigure}
 \begin{subfigure}{0.15\textwidth}
 \includegraphics[width=1\textwidth]{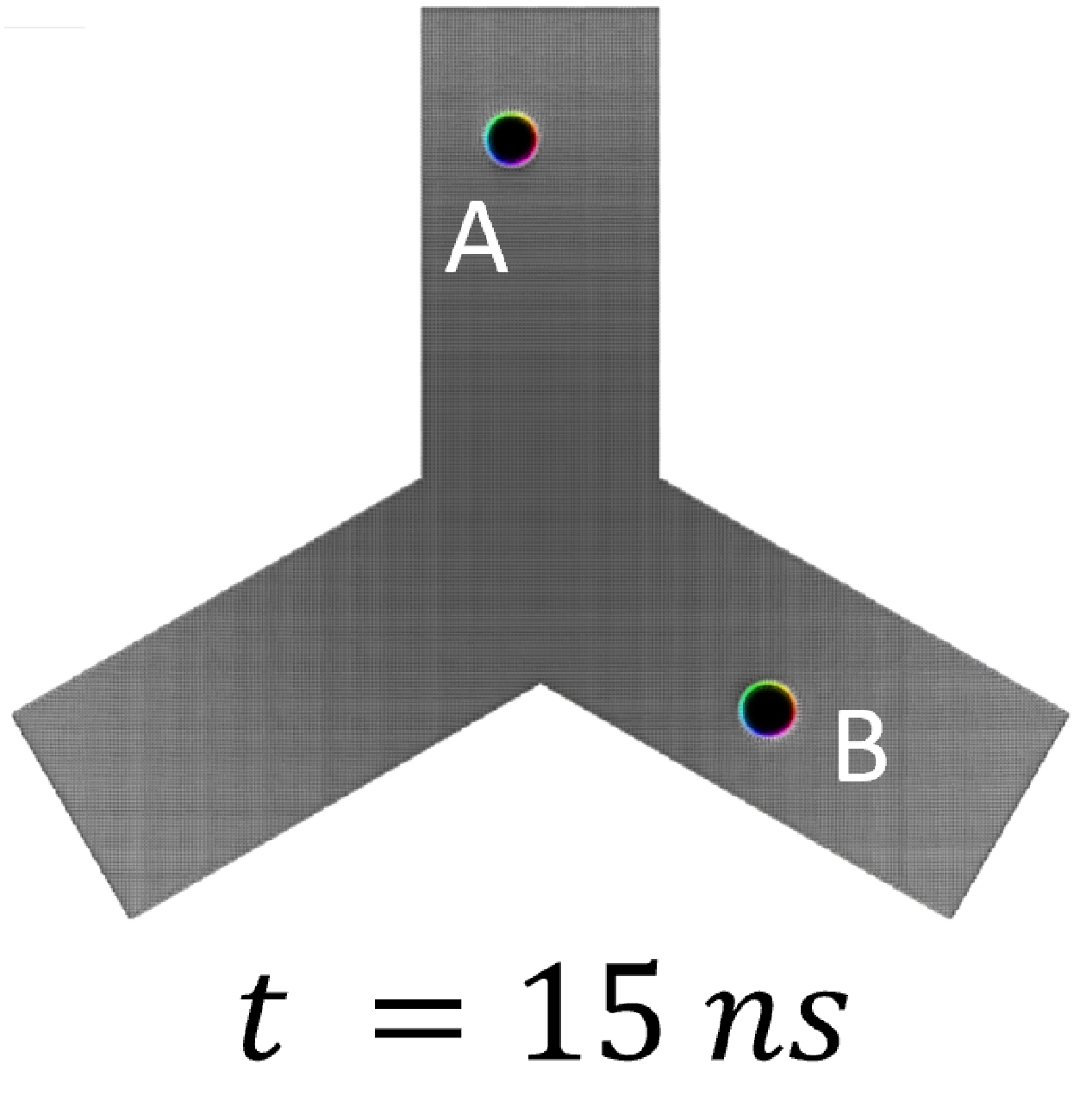}
 \caption{\label{fig:t2}}
 \end{subfigure}
 \begin{subfigure}{0.15\textwidth}
 \includegraphics[width=1\textwidth]{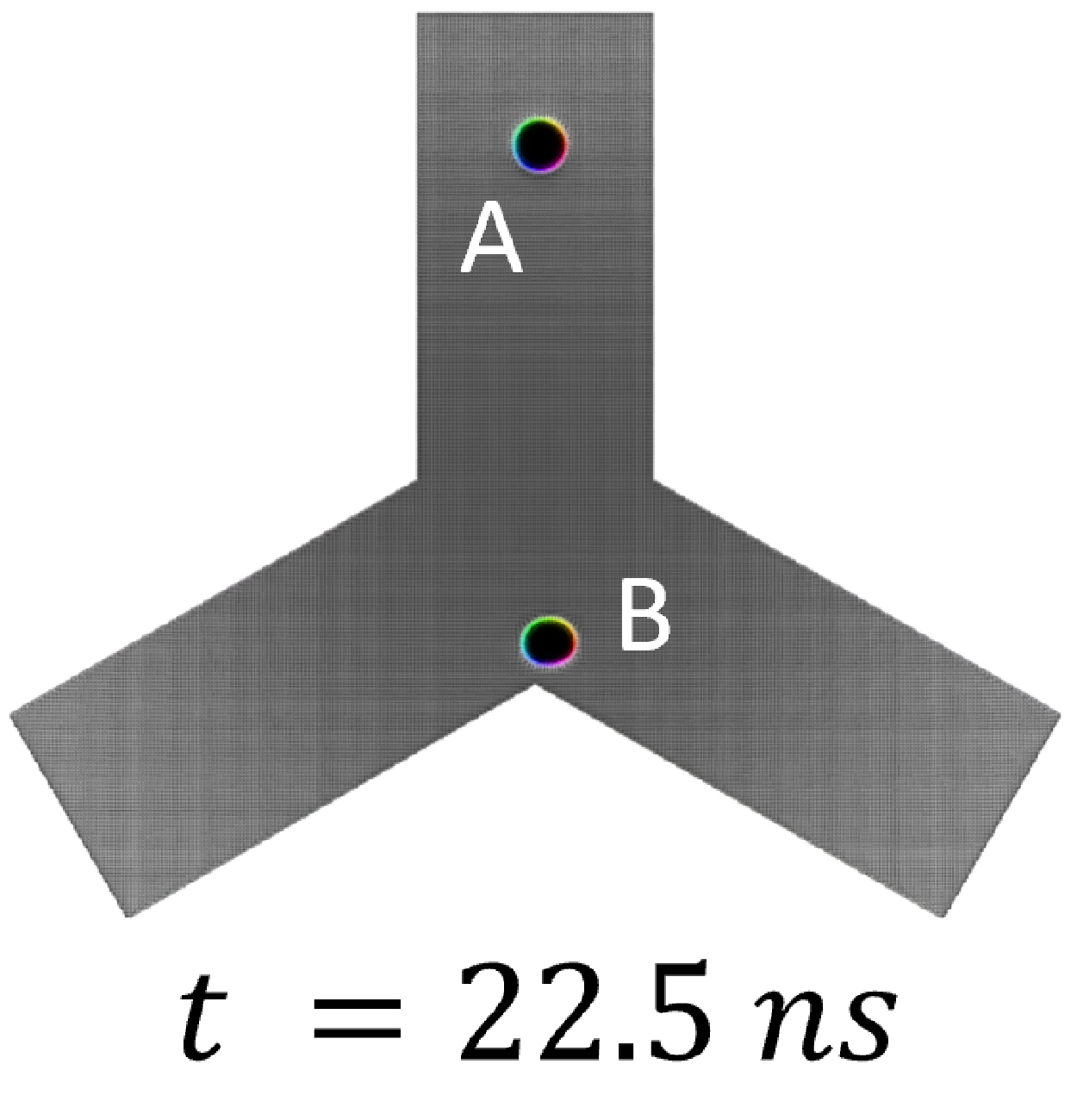}
 \caption{\label{fig:t3}}
 \end{subfigure}
 \begin{subfigure}{0.15\textwidth}
 \includegraphics[width=1\textwidth]{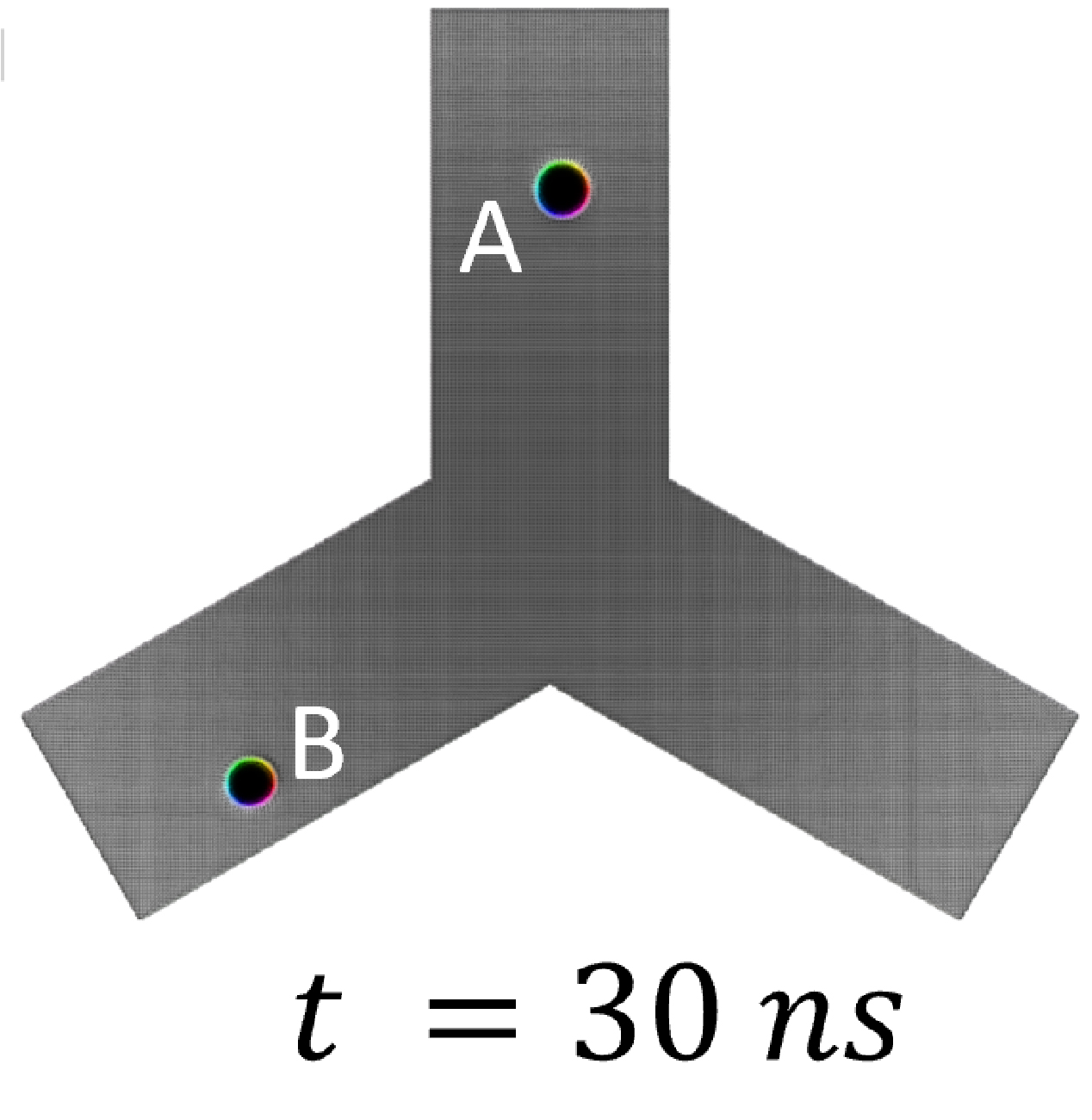}
 \caption{\label{fig:t4}}
 \end{subfigure}
 \begin{subfigure}{0.15\textwidth}
 \includegraphics[width=1\textwidth]{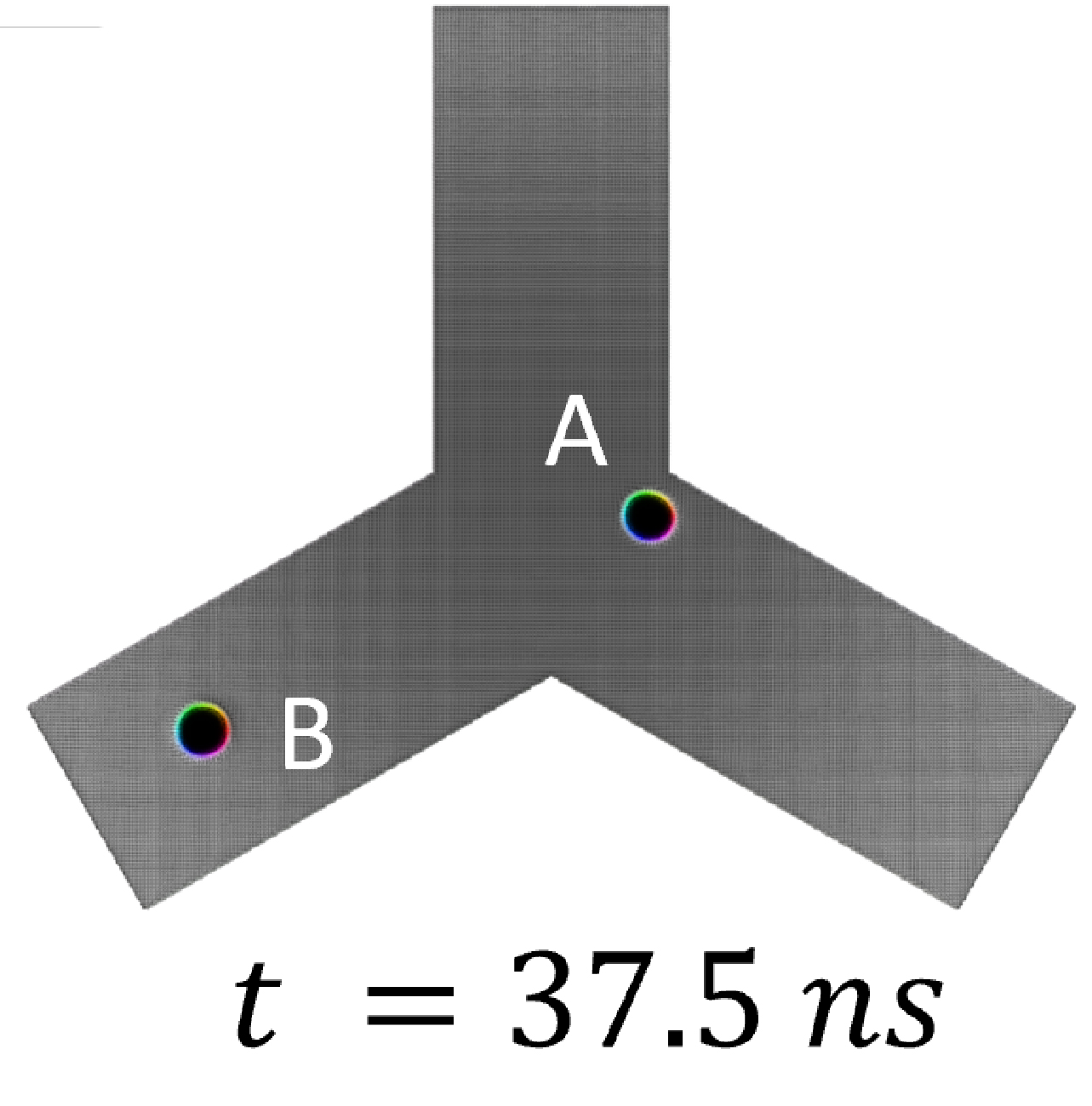}
 \caption{\label{fig:t5}}
 \end{subfigure}
 \begin{subfigure}{0.15\textwidth}
 \includegraphics[width=1\textwidth]{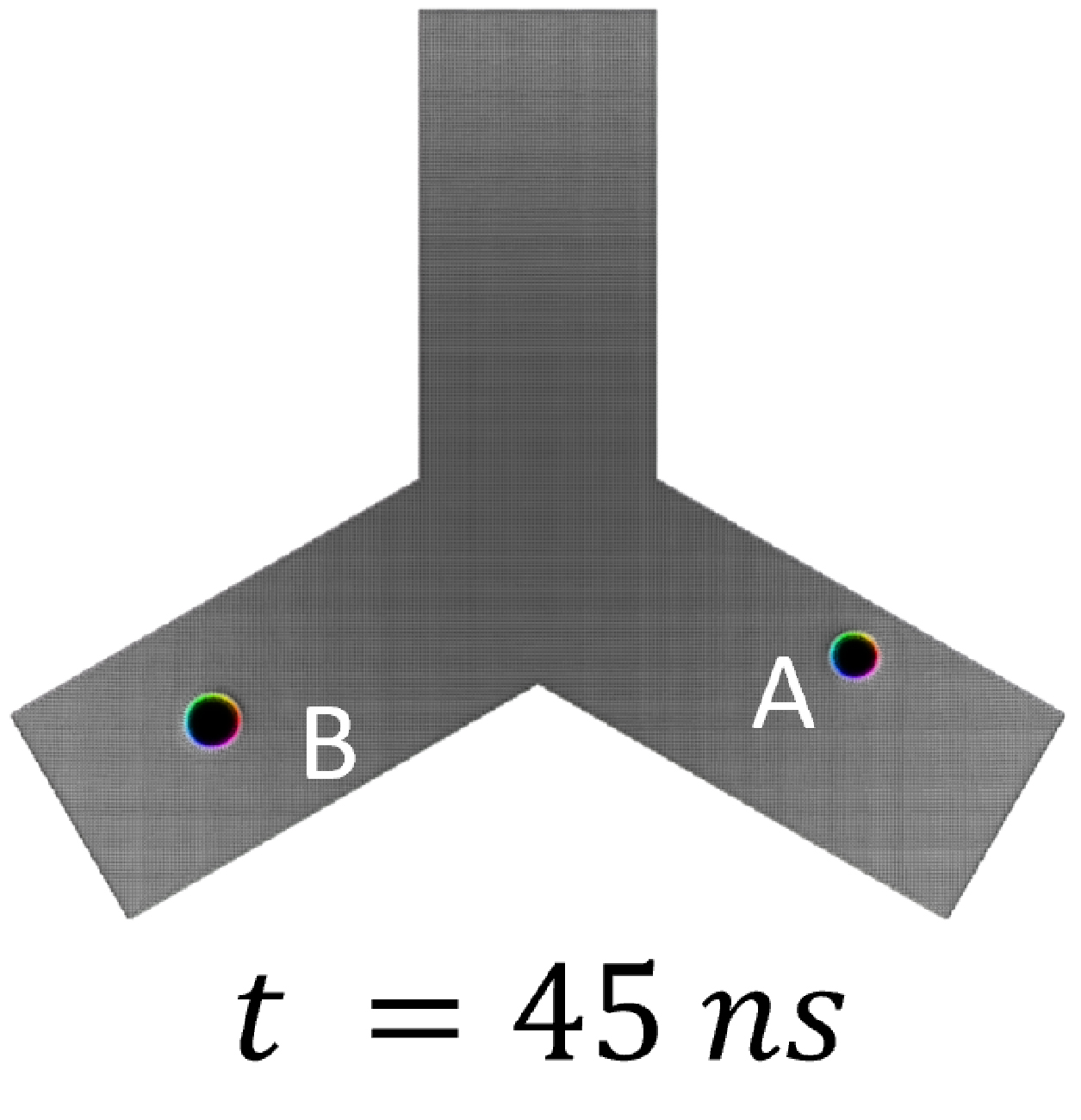}
 \caption{\label{fig:t6}}
 \end{subfigure}
\caption{\label{fig:two} (a) Schematic of our braiding process: manipulations of four skyrmions in the MML track are shown. MBS at the centers of vortices bound to each of these skyrmions are labeled $\gamma_1$--$\gamma_4$. Ohmic contacts in HM layers of the MML are shown in brown and rf readout lines are shown in green. II--VI show the steps involved in braiding $\gamma_2$ and $\gamma_4$. In step II, $\gamma_1$ and $\gamma_2$ are brought close to rf lines by applying charge currents from C to A and D to B, respectively. $\gamma_1$ and $\gamma_2$ are then initialized by performing a dispersive readout of their parity (see Section~\ref{sec:read}). Similarly, $\gamma_3$ and $\gamma_4$ are initialized after applying charge currents along P to R and Q to S, respectively. In step III, $\gamma_2$ is moved aside to make room for $\gamma_4$ by applying currents from B to X followed by applying currents from X to C. In step IV, $\gamma_4$ is braided with $\gamma_2$ by applying currents along S to X and X to B. Finally, in step V, the braiding process is completed by bringing $\gamma_2$ to S by applying currents from A to X and from X to S. Parities (i.e., fusion outcomes) of $\gamma_1$ and $\gamma_4$, and $\gamma_3$ and $\gamma_2$ are then measured in step VI. Fusion outcomes in each pair of MBS indicate the presence or absence of a fermion corresponding to a parity of $\pm1$ \cite{PhysRevApplied.12.054035, PhysRevX.6.031016}. (b) Initial position of the skyrmions labeled A and B in the micromagnetic simulation for skyrmion braiding (see Appendix.~\ref{app:A}) (c--h) Positions of the two skyrmions at the given times as the braiding progresses. Charge current $j = 2\times 10^{12}$ A/m$^2$ was applied.}

\end{figure*}

Our setup consists of a thin TSC layer that hosts vortices grown on top of a MML that hosts skyrmions as shown in Fig.~\ref{fig:schematic}. A thin insulating layer separates the magnetic and superconducting layers ensuring electrical separation between the two. Vortices in a TSC are expected to host MBS at their cores \cite{wang2018evidence,chen2020observation, chen2018discrete}. Stray fields from a skyrmion in the MML nucleate such a vortex in the TSC, forming a bound skyrmion--vortex pair under favorable energy conditions (see Sec.~\ref{sec:initial}). This phenomenon has been recently experimentally demonstrated in Ref.~\cite{petrovic2021skyrmion}, where stray fields from N\'eel skyrmions in Ir/Fe/Co/Ni magnetic multilayers nucleated vortices in a bare Niobium superconducting film.

The MML consists of alternating magnetic and heavy metal (HM) layers, as shown in Fig.~\ref{fig:layers}. The size of a skyrmion in a MML is determined by a delicate balance between exchange, magnetostatic, anisotropy and Dzyaloshinskii–Moriya interaction (DMI) energies \cite{wang2018theory, romming2015field} -- and the balance is highly tunable, thanks to advances in spintronics \cite{buttner2018theory, dupe2016engineering, soumyanarayanan2017tunable}. Given a TSC, this tunability allows us to find a variety of magnetic materials and skyrmion sizes that can satisfy the vortex nucleation condition [to be detailed in Eq.~(\ref{eqn:nuc})]. In Appendix~\ref{app:A}, we provide a specific example of FeTeSe topological superconductor coupled with Ir/Fe/Co/Ni magnetic multilayers.

Due to large intrinsic spin-orbit coupling, a charge current through the heavy metal layers of a MML exerts spin-orbit torques (SOT) on the magnetic moments in the MML, which have been shown to drive skyrmions along magnetic tracks \cite{fert2013skyrmions, woo2017spin}. In our platform, to realize Majorana braiding we propose to pattern the MML into a track as shown in Fig.~\ref{fig:schematic} and use local spin-orbit torques to move skyrmions along each leg of the track. If skyrmions are braided on the MML track, and if skyrmion-vortex binding force is stronger than total pinning force on the SVPs, then the MBS hosting vortices in TSC will closely follow the motion of skyrmions, resulting in the braiding of MBS. We note here that there is an upper threshold speed with which a SVP can be moved as detailed in Sec.~\ref{sec:braid}. By using experimentally-relevant parameters for TSC and MML in Appendix~\ref{app:A}, we show that our Majorana braiding scheme can be realized with existing materials.

We propose a non-demolition microwave measurement technique for the readout of the quantum information encoded in a pair of vortex Majorana bound states (MBS). A similar method has been proposed for the parity readout in topological Josephson junctions~\cite{PhysRevB.92.245432,Vayrynen2015,Yavilberg2015,PhysRevB.99.235420,PRXQuantum.1.020313} and in Coulomb blockaded Majorana islands~\cite{PhysRevB.95.235305}. Dipole moments of transitions from MBS to CdGM levels couple dispersively to electric fields in a microwave cavity, producing a parity-dependent dispersive shift in the cavity resonator frequency. Thus by probing the change in the resonator's natural frequency, the state of the Majorana modes can be inferred. Virtual transitions from Majorana subspace to excited CdGM subspace induced due to coupling to the cavity electric field are truly parity conserving, making our readout scheme a so-called topological quantum non-demolition technique \cite{PRXQuantum.1.020313, PhysRevB.99.235420}. The readout scheme is explained in greater detail in Sec.~\ref{sec:read}.

As discussed above, in our platform we consider coupling between a thin superconducting layer and magnetic multilayers. We note that in thin superconducting films, vortices are characterized by the Pearl penetration depth, given by $\Lambda \ =\ \lambda ^{2} /d_{s}$, where $\lambda$ is the London penetration depth and $d_{s}$ is the thickness of the TSC film. Typically, these penetration depths $\Lambda$ are much larger than skyrmion radii $r_{sk}$ in MMLs of interest. Further, interfacial DMI in MML stabilizes a N\'eel skyrmion as opposed to a Bloch skyrmion. So hereon, we only study coupling between a N\'eel skyrmion and a Pearl vortex in the limit $\Lambda\gg r_{sk}$.

\section{\label{sec:initial}Initialization and SVP in Equilibrium}

Fig.~\ref{fig:flow} illustrates the process flow of our initialization scheme. Skyrmions can be generated individually in MML by locally modifying magnetic anisotropy through an artificially created defect center and applying a current through adjacent heavy metal layers \cite{zhang2020skyrmion}. Such defect centers have been experimentally observed to act as skyrmion creation sites \cite{buttner2017field}. When the TSC--MML heterostructure is cooled below the superconducting transition temperature (SC $T_{C}$), stray fields from a skyrmion in the MML will nucleate a vortex and an antivortex in the superconducting layer if the nucleation leads to a lowering in overall free energy of the system \cite{volkov}. An analytical expression has been obtained for the nucleation condition in Ref.~\cite{NeelInteraction} ignoring contributions of dipolar and Zeeman energies to total magnetic energy: a N\'eel skyrmion nucleates a vortex directly on top of it if 
\begin{equation}
    d_{m}\left[ \alpha _{K}\frac{Kr_{sk}^{2}}{2} -\alpha _{A} A-M_{0} \phi _{0}\right] \geq \frac{{\phi _{0}}^2}{8 \pi^2 \lambda} \ln\left(\frac{\Lambda }{\xi }\right).
    \label{eqn:nuc}
\end{equation}
\noindent Here, $d_{m}$ is the effective thickness, $M_{0}$ is the saturation magnetization, $A$ is the exchange stiffness and $K$ is the perpendicular anisotropy constant of the MML; $\alpha_K$ and $\alpha_A$ are positive constants that depend on skyrmion's spatial profile (see Appendix~\ref{app:A}), $r_{sk}$ is the radius of the skyrmion in the presence of a Pearl vortex \footnote{The radius of a skyrmion is not expected to change significantly in the presence of a vortex \cite{NeelInteraction}. We verified this claim with micromagnetic simulations. For the materials in Appendix~\ref{app:A}, when vortex fields are applied on a bare skyrmion, its radius increased by less than $10\%$. So, for numerical calculations in this paper, we use bare skyrmion radius for $r_{sk}$.}, $\phi _{0}$ is the magnetic flux quantum, and $\Lambda$ ($\xi$) is the Pearl depth (coherence length) of the TSC. Although a complete solution of the nucleation condition must include contributions from dipolar and Zeeman energies to total energy of a MML, such a calculation can only be done numerically and Eq.~(\ref{eqn:nuc}) can still be used as an approximate estimate. For the choice of materials listed in the Appendix, the left side of the equation exceeds the right side by $400\%$, strongly suggesting the nucleation of a vortex for every skyrmion in the MML. Furthermore, skyrmions in Ir/Fe/Co/Ni heterostructures have also been experimentally shown to nucleate vortices in Niobium superconducting films \cite{petrovic2021skyrmion}. 

We proceed to characterize the strength of a skyrmion (Sk) -- vortex (Vx) binding force as it plays a crucial role in determining the feasibility of moving the skyrmion and the vortex as a single object. Spatial magnetic profile of a N\'eel skyrmion is given by $\boldsymbol{M}_{sk} =M_{0}[\zeta \sin\theta(r) \boldsymbol{\hat{r}}+ \cos\theta(r) \boldsymbol{\hat{z}}]$, where $\zeta=\pm$1 is the chirality and $\theta(r)$ is the angle of the skyrmion. For $\Lambda\gg r_{sk}$,  the interaction energy between a vortex and a skyrmion is given by \cite{NeelInteraction}:
\begin{equation}
    E_{Sk-Vx} =\frac{M_{0} \phi _{0} r_{sk}^{2}}{2\Lambda }\int_{0}^{\infty} \frac{1}{q^2}(e^{-q\tilde{d}}-1) J_{0}(qR) m_{z,\theta}(q) \,dq,
    \label{eqn:energy}
\end{equation}

\noindent where $\tilde{d} = d_m \slash r_{sk}$, $J_{n}$ is the nth-order Bessel function of the first kind, and $R=r/r_{sk}$ is the normalized horizontal displacement $r$ between the centers of the skyrmion and the vortex. $m_{z,\theta}(q)$ contains information about skyrmion's spatial profile and is given by \cite{NeelInteraction}: $m_{z,\theta}(q) = \int_{0}^{\infty} x [\zeta q + \theta^\prime ( x )] J_{1}( qx) \sin\theta(x) \,dx$, where $\theta ( x )$ is determined by skyrmion ansatz.

We now derive an expression for the skyrmion--vortex restoring force by differentiating Eq.~(\ref{eqn:energy}) with respect to $r$:
\begin{equation}
    F_{Sk-Vx} =-\frac{M_{0} \phi _{0} r_{sk}}{2\Lambda }\int_{0}^{\infty} \frac{1}{q}(1- e^{-q\tilde{d}}) J_{1}(qR) m_{z,\theta}(q) \,dq.
    \label{eqn:force}
\end{equation}
For small horizontal displacements $r\ll r_{sk}$ between the centers of the skyrmion and the vortex, we can approximate the Sk--Vx energy as:
\begin{equation}
    E_{Sk-Vx} =\frac{1}{2} kr^{2},
    \label{eqn:springconstant}
\end{equation}
\noindent with an effective spring constant 
\begin{equation}
    k =-\frac{M_{0} \phi _{0}}{4\Lambda }\int_{0}^{\infty} (1- e^{-q\tilde{d}}) m_{z,\theta}(q) \,dq.
    \label{eqn:spring}
\end{equation}

Figs.~\ref{fig:energies}--\ref{fig:forces} show binding energy and restoring force between a vortex and skyrmions of varying thickness for the materials listed in Appendix~\ref{app:A}. Here we used domain wall ansatz for the skyrmion with $\theta(x) = 2\tan^{-1}[\frac{\sinh(r_{sk}/\delta)}{\sinh(r_{sk}x/\delta)}]$, where $r_{sk}/\delta$ is the ratio of skyrmion radius to its domain wall width and $x$ is the distance from the center of the skyrmion normalized by $r_{sk}$. As seen in Fig.~\ref{fig:forces}, the restoring force between a skyrmion and a vortex increases with increasing separation between their centers until it reaches a maximum value, $F_{max}$, and then decreases with further increase in separation. We note that $F_{max}$ occurs when Sk--Vx separation is equal to the radius of the skyrmion, i.e. when $R=1$ in Eq.~(\ref{eqn:force}):
\begin{equation}
    F_{max} = -\frac{M_{0} \phi _{0} r_{sk}}{2\Lambda }\int_{0}^{\infty} \frac{1}{q}(1- e^{-q\tilde{d}}) J_{1}(q) m_{z,\theta}(q) \,dq. 
    \label{eqn:fmax}
\end{equation}

\noindent As the size of the skyrmion increases, the maximum binding force $F_{max}$ of the SVP increases. For a given skyrmion size, increasing the skyrmion thickness increases the attractive force until the thickness reaches the size of the skyrmion. Further increase in MML thickness does not lead to an appreciable increase in stray fields outside the MML layer and, as a result, the Sk--Vx force saturates.

It is important to note that stray fields from a skyrmion nucleate both a vortex and an antivortex (Avx) in the superconducting layer \cite{volkov, PhysRevLett.88.017001, milosevic_guided_2010, PhysRevLett.93.267006}. While the skyrmion attracts the vortex, it repels the antivortex. Eqs.~(\ref{eqn:energy}) and (\ref{eqn:force}) remain valid for Sk--Avx interaction, but switch signs. The equilibrium position of the antivortex is at the location where repulsive skyrmion--antivortex force, $F_{Sk-Avx}$, is balanced by the attractive vortex--antivortex force, $F_{Vx-Avx}$~\cite{lemberger2013theory, ge2017controlled}. Fig.~\ref{fig:fvav} shows $F_{Vx-Avx}$ against $F_{Sk-Avx}$ for the platform in the Appendix. We see that for thicker magnets, the location of the antivortex is far away from that of the vortex, where the Avx can be pinned with artificially implanted pinning centers \cite{aichner2019ultradense, gonzalez2018vortex}. For thin magnetic films, where the antivortex is expected to be nucleated right outside the skyrmion radius, we can leverage Berezinskii–Kosterlitz–Thouless (BKT) transition to negate $F_{Vx-AVx}$ for Vx-Avx distances $r<\Lambda$ \cite{PhysRevB.104.024509, schneider_excess_2014, goldman2013berezinskii, zhao2013evidence}. Namely, when a Pearl superconducting film is cooled to a temperature below $T_C$ but above $T_{BKT}$, vortices and antivortices dissociate to gain entropy, which minimizes the overall free energy of the system \cite{beasley1979possibility}. While the attractive force between a vortex and an antivortex is nullified, a skyrmion in the MML still attracts the vortex and pushes the antivortex towards the edge of the sample, where it can be pinned. Therefore we assume that the antivortices are located far away and neglect their presence in our braiding and readout schemes.

\section{\label{sec:braid}Braiding}

Majorana braiding statistics can be probed by braiding a pair of MBS \cite{RevModPhys.80.1083} which involves swapping positions of the two vortices hosting the MBS. We propose to pattern the MML into interconnected Y-junctions as shown in Fig.~\ref{fig:two} to enable that swapping. Ohmic contacts in HM layers across each leg of the Y-junctions enable independent application of charge currents along each leg of the track. These charge currents in-turn apply spin-orbit torques on the adjacent magnetic layers and enable skyrmions to be moved independently along each leg of the track. As long as skyrmion and vortex move as a collective object, braiding of skyrmions in the MML leads to braiding of MBS hosting vortices in the superconducting layer. Below we study the dynamics of a SVP subjected to spin torques for braiding. We calculate all external forces acting on the SVP in the process and discuss the limits in which the skyrmion and the vortex move as a collective object.

For a charge current $\bm{J}$ in the HM layer, the dynamics in the magnetic layer is given by the modified Landau–Lifshitz–Gilbert (LLG) equation \cite{hayashi2014quantitative, slonczewski1996current}:
\begin{equation}
    \partial _{t}\bm{m} =-\gamma (\bm{m} \times {{\bm H}_{eff}} +\eta J\ \bm{m} \times \bm{m} \times \bm{p}) +\alpha \bm{m} \times \partial _{t}\bm{m}
    \label{eqn:llg}
\end{equation}
\noindent where we have included damping-like term from the SOT and neglected the field-like term as it does not induce motion of N\'eel skyrmions for our geometry \cite{jiang_blowing_2015}. Here, $\gamma$ is the gyromagnetic ratio, $\alpha$ is the Gilbert damping parameter, and ${{\bm H}_{eff}}$ is the effective field from dipole, exchange, anisotropy and DMI interactions. $\bm{p}=sgn(\Theta _{SH})\bm{\hat{J}} \times \hat{\bm{n}}$ is the direction of polarization of the spin current, where $\Theta _{SH}$ is the spin Hall angle, $\bm{\hat{J}}$ is the direction of charge current in the HM layer and $\hat{\bm{n}}$ is the unit vector normal to the MML. $\eta=\hbar \Theta _{SH}/2eM_{0} d_{m}$ quantifies the strength of the torque, $\hbar$ is the reduced Planck's constant and $e$ is the charge of an electron. 

Assuming skyrmion and vortex move as a collective object, semiclassical equations of motion for the centers of mass of the skyrmion and the vortex can be written using collective coordinate approach as done in Ref.~\cite{hals2016composite}:
\begin{eqnarray}
    m_{sk}\ddot{\bm{R}}_{sk}= {\bf{F}}_{SOT} - \frac{\partial U_{sk,\ pin}}{\partial \bm{R}_{sk}} - & {\bm{G}}_{sk}\times \dot{\bm{R}}_{sk} - 4\pi s \alpha \dot{\bm{R}}_{sk} \nonumber \\
    &- k({\bm{R}}_{sk}-{\bm{r}}_{vx}),
    \label{eqn:skmotion}
\end{eqnarray}
and
\begin{eqnarray}
    m_{vx}\ddot{\bm{R}}_{vx} = - \frac{\partial U_{vx,\ pin}}{\partial \bm{R}_{vx}} - &{\bm{G}}_{vx}\times \dot{\bm{R}}_{vx} - {\alpha}_{vx} \dot{\bm{R}}_{vx} \nonumber \\
    & + k({\bm{R}}_{sk}-{\bm{r}}_{vx}),
    \label{eqn:vxmotion}
\end{eqnarray}
\noindent where ${\bm{R}}_{sk}$ (${\bm{R}}_{vx}$), $m_{sk}$ ($m_{vx}$) and $q_{sk}$ ($q_{vx}$)  are the position, mass and chirality of the skyrmion (vortex). $k$ is the effective spring constant of the Sk--Vx system, given in Eq.~(\ref{eqn:spring}). ${\bm{F}}_{SOT}=\pi ^{2} \gamma \eta r_{sk} s\bm{{J}} \times \hat{\bm{n}}$ is the force on a skyrmion due to spin torques in Thiele formalism, where $s=M_0 d_m/\gamma$ is the spin density \cite{upadhyaya2015electric, thiele1970theory}. The third term on the right side of Eq.~(\ref{eqn:skmotion}) gives Magnus force on the skyrmion, with ${\bm{G}}_{sk} = 4\pi s q_{sk}\hat{\bm{z}}$, and the fourth term characterizes a dissipative force due to Gilbert damping.  Similarly, the second term on the right side of Eq.~(\ref{eqn:vxmotion}) gives the Magnus force on the vortex with ${\bm{G}}_{vx} = 2\pi s n_{vx} q_{vx} \hat{\bm{z}}$, with $n_{vx}$ being the superfluid density of the TSC, and the third term characterizes viscous force with friction coefficient ${\alpha}_{vx}$. $U_{sk,\ pin}$ ($U_{vx,\ pin}$) gives the pinning potential landscape for the skyrmion (vortex). The last term in Eq.~(\ref{eqn:vxmotion}) represents restoring force on a vortex due to its separation from a skyrmion and is valid when $\mid{\bm{R}}_{sk}-{\bm{R}}_{vx}\mid <r_{sk}$. Here, $k$ is the effective spring constant characterizing Sk--Vx force, as given by Eq.~(\ref{eqn:springconstant}).

We consider steady-state solutions of the equations of motion assuming that the skyrmion and the vortex are bound. We discuss conditions for the dissociation of a SVP later. For a given external current $\bm{J}$, velocity $v$ of a SVP in steady state is obtained by setting $\ddot{\bm{R}}_{sk} = \ddot{\bm{R}}_{vx} = 0$ and $\dot{\bm{R}}_{sk} = \dot{\bm{R}}_{vx} = \dot{\bm{R}}$ in Eqs.~(\ref{eqn:skmotion}) and (\ref{eqn:vxmotion}):
\begin{equation}
    v = |\dot{\bm{R}}| = \frac{\pi ^{2} \gamma \eta r_{sk} sJ}{\sqrt{(G_{sk}+G_{vx})^{2} +(4\pi s \alpha + \alpha_{vx})^{2}}}.
    \label{eqn:vgivenj}
\end{equation}

\noindent In general, the SVP moves at an angle $\varphi$ relative to $\bm{{F}}_{SOT}$ due to Magnus forces on the skyrmion and the vortex, with:
\begin{eqnarray}
    \tan \varphi = \frac{G_{sk}+G_{vx}}{4\pi s \alpha + \alpha_{vx}}.
    \label{eqn:svpangle}
\end{eqnarray}

Armed with the above equations, we extract some key parameters that determine the feasibility of our braiding scheme. First, if $\bm{{F}}_{SOT}$ from external currents is unable to overcome the maximum pinning force on either the skyrmion ($F_{pin, sk}$) or the vortex ($F_{pin, vx}$), the SVP will remain stationary. This gives us a lower threshold $J^-$ on the external current which is obtained by weighing $\bm{{F}}_{SOT}$ against the pinning forces:
\begin{equation}
    J^{-} = \frac{max(F_{pin, sk}, F_{pin, vx})}{\pi ^{2} \gamma \eta r_{sk} s}.
    \label{eqn:jminus}
\end{equation}
Second, once the SVP is in motion, drag and Magnus forces that act on the skyrmion and the vortex are proportionate to their velocity. If the net external force on a vortex in motion is larger than the maximum force with which a skyrmion can pull it ($F_{max}$), then the skyrmion and the vortex dissociate and no longer move as a collective object. This sets an upper bound $v^+$ on the SVP speed which can be obtained by balancing $F_{max}$ with the net force from Magnus and drag forces on the vortex. This maximum speed  plays a key role in determining whether our braiding and readout scheme can be completed within the quasiparticle poisoning time.

\begin{equation}
    v^{+} = \frac{F_{max}}{\sqrt{(\alpha_{vx})^2+(G_{vx})^2}}.
    \label{eqn:vplus}
\end{equation}
An upper bound on the SVP speed implies an upper bound $J^+$ on the external current which can be obtained by putting $v^+$ in Eq.~(\ref{eqn:vgivenj}):
\begin{equation}
    J^{+} = \frac{v^{+} \sqrt{(G_{sk}+G_{vx})^{2} +(4\pi s \alpha + \alpha_{vx})^{2}}}{\pi ^{2} \gamma \eta r_{sk} s}.
    \label{eqn:jplus}
\end{equation}
Another parameter of critical importance, the distance of closest approach between two skyrmion--vortex pairs ($r_{min}$) plays a crucial role in achieving significant overlap of the MBS wavefunctions centered at the vortex cores and is given by balancing Sk--Vx attractive force by Vx--Vx repulsive force:
\begin{equation}
    r_{min} = \frac{\phi_0^2}{4\pi^2 \Lambda} \frac{1}{F_{max}}.
    \label{eqn:rmin}
\end{equation}
Finally, the power $P$ dissipated in heavy metal layers due to Joule heating from charge currents has to be effectively balanced by the cooling rate of the dilution refrigerator: 
\begin{equation}
    P = n_{hm} L W t_{hm}  \rho_{hm} J^2,
    \label{eqn:power}
\end{equation}
\noindent where $n_{hm}$ is the number of heavy metal layers, $L$ ($W$) is the length (width) of the active segment of the MML track, $t_{hm}$ is the thickness of each heavy metal layer and $\rho_{hm}$ is the resistivity of a heavy metal layer.

By applying a current $J^- < J < J^+$ locally in a desired section of the MML track, each SVP can be individually addressed. For the materials listed in Appendix~\ref{app:A}, the maximum speed $v^+$ with which a SVP can be moved is over $1000$ m/s. At this top speed, SVPs can cover the braiding distance (the sum of the lengths of the track in steps I--VI of Fig.~\ref{fig:braiding}) of $50 r_{sk}$ in about $0.15$ ns, but the process generates substantial Joule heating. At a reduced speed of $0.25$ m/s, SVPs cover that distance in $7$ $\mu$s generating $30$ $\mu$W of heat during the process, which is within the cooling power of modern dilution refrigerators. SVPs can be braided at faster speeds if the dilution fridges can provide higher cooling power or if the resistivity of heavy metal layers in the MML can be lowered. Although quasiparticle poisoning times in superconducting vortices have not been measured yet, estimates in similar systems range from hundreds of microseconds to seconds \cite{higginbotham2015parity, PhysRevLett.126.057702, PhysRevB.85.174533}. Our braiding time falls well within such estimates for quasiparticle poisoning times, indicating the viability of our platform. Furthermore, the ability to easily tune braiding times in our platform by varying magnitude of currents in heavy metal layers can be used to investigate the effects of quasiparticle poisoning on the braiding protocol.

As will be shown in Section~\ref{sec:read}, Vx--Vx distances $<10\xi$ should be sufficient to perform a dispersive readout of MBS parity in adjacent vortices. For the materials listed in Appendix~\ref{app:A}, the distance of closest approach between two vortices is $r_{min}=40$ nm. The shape of the MML track further limits how close two vortices can be brought together (see step II in Fig.~\ref{fig:braiding}). With the geometry of the track taken into account, Vx--Vx distance less than $10\xi$ can still be easily achieved, enough to induce a detectable shift in cavity's resonance frequency during the dispersive readout.

Figs.~\ref{fig:t0}--\ref{fig:t6} show the results of micromagnetic simulation of braiding skyrmions in a smaller section of the MML (for computational reasons) for the example platform. The details of the simulation are given in the Appendix~\ref{app:A}. The simulation results demonstrate the effectiveness of using local SOT to move individual skyrmions and realize braiding. Finally, as discussed in this section, due to the strong skyrmion--vortex binding force, vortices hosting MBS in the TSC will braid alongside the skyrmions.

\begin{figure*}
\centering
 \includegraphics[width=1\textwidth]{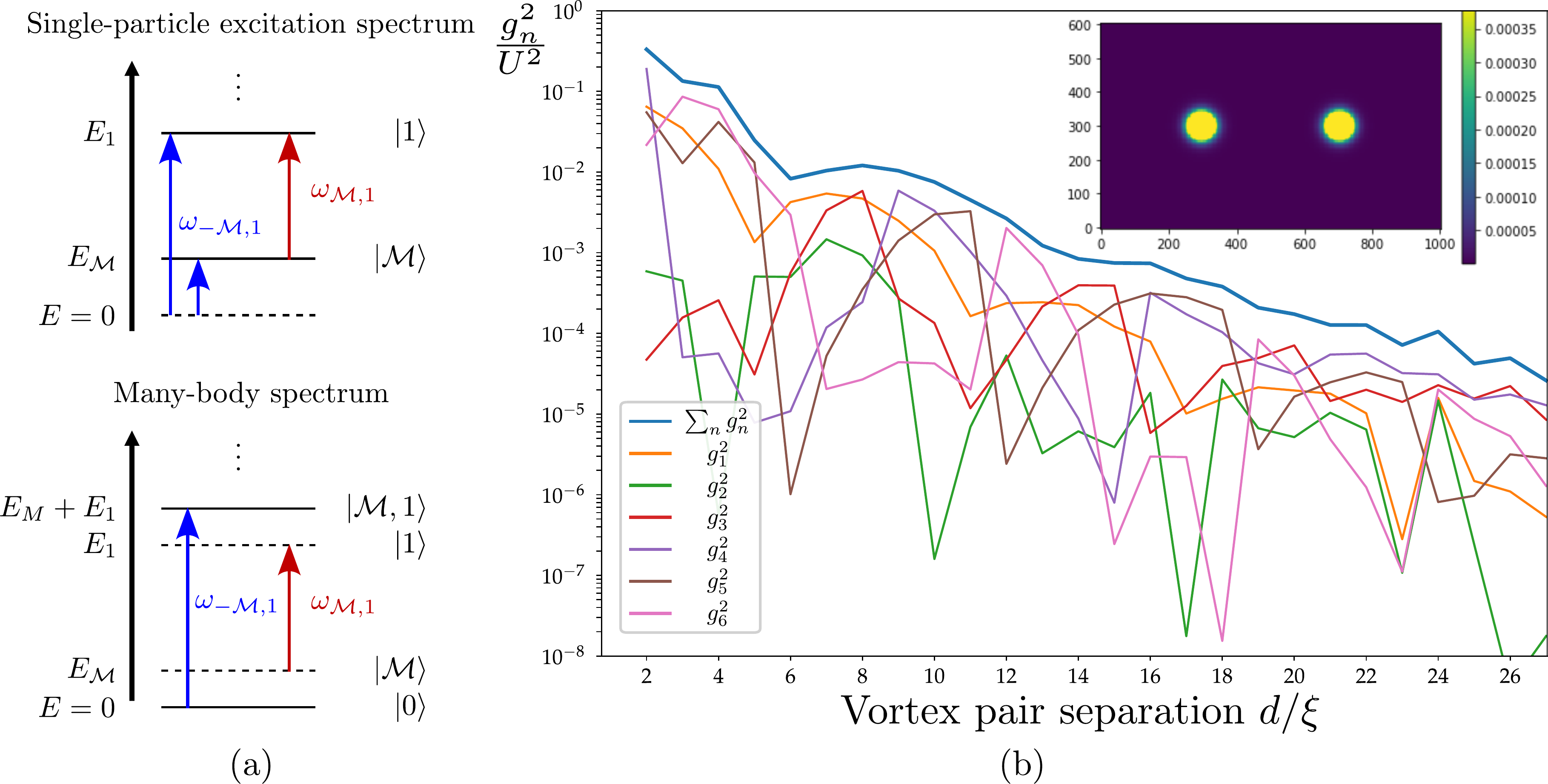}
 \caption{\label{fig:readout} (a) Schematic of our readout process. When two vortices are brought close, microwave transitions can be dispersively driven from the MBS to the excited hybridized CdGM levels (only level $1$ is shown). Parity of the Majorana mode can be inferred from the difference in the cavity frequency shift produced by $\omega_{-\mathcal{M},1}$ and $\omega_{\mathcal{M},1}$ transitions (see Eq.~(\ref{eq:chi})). 
 The allowed fermion parity conserving transitions are shown in both single-particle and many-particle representations. In the latter, dashed and solid lines denote states in the two fermion parity sectors. 
 The transition of frequency $\omega_{-\mathcal{M},1}$ (blue arrows) corresponds to breaking a Cooper pair and exciting the MZM and CdGM levels (MZM being initially unoccupied). 
 When the MZM is occupied, the transition of frequency $\omega_{\mathcal{M},1}$ (red arrow) excites the MZM quasiparticle  into the CdGM level. 
 The dipole transition matrix elements are different for the two processes, enabling parity readout. 
 (b) MZM-parity sensitive dipole transition strength versus vortex pair separation. We denote $g^2_n = (|\mathbf{E}_{0}\cdot\mathbf{d}_{n,- \mathcal{M} }|^{2}-|\mathbf{E}_{0}\cdot\mathbf{d}_{n, \mathcal{M} }|^{2}) $ the dipole transition strength between the Majorana level and the $n$th CdGM level. We plot the dimensionless strength normalized by $U = e |\mathbf{E}_{0}| l$. 
As expected from MZM hybridization, $g_n$ decays approximately exponentially in the distance between the two vortices. Oscillations in $g^2_n$ represent oscillations in the wave functions of a clean system. In a disordered (real) system the oscillations are expected to be smeared out.
The inset shows the probability density for the MZM hosted by a vortex pair 400 nm apart. 
The simulation was done for an effective 2D model (a $1000\times 600 \mathrm{nm}^2$ rectangle) of a 3D topological insulator surface, see Refs.~\cite{PhysRevB.86.155146,PhysRevX.7.031006,MW_inprep}.  We used $\xi = 15 \mathrm{nm}$, vortex radius $r = \xi$, and $E_F = 1.125 \Delta$ in the vortex.  
}

\end{figure*}

\section{\label{sec:read}Readout}

Quantum information is encoded in the charge parity of MBS hosted in a pair of vortices which we propose to readout with a dispersive measurement technique. Fig.~\ref{fig:readout}a summarizes our readout scheme - the top figure shows single particle energy levels and the bottom figure shows many-body energy levels of a pair of vortices brought close to each other. In the dispersive limit, a microwave cavity electric field can drive virtual transitions from the ground state Majorana manifold to excited CdGM manifold (only one CdGM level, labeled 1, is considered in the figure). The transitions allowed by selection rules, labeled $\omega_{-\mathcal{M},1}$ and $\omega_{\mathcal{M},1}$ are shown in the many-body spectrum. Each of these virtual transitions causes a state-dependent dispersive shift in the cavity's natural frequency and the parity of the vortex pair can be inferred from relative change in cavity frequency. Note that each of the allowed transitions is truly parity conserving since microwaves cannot change the number of fermions. 
Since parity states are true eigenstates (as opposed to approximate eigenstates) of the measurement operation, our readout scheme can be dubbed as a topological quantum non-demolition technique \cite{PRXQuantum.1.020313, PhysRevB.99.235420}. We now proceed to calculate dipole coupling strengths of the allowed transitions to cavity electric field and the corresponding dispersive shift.

In BCS mean-field theory, the coupling to electric field can be described by the  Hamiltonian 
\begin{equation}
\delta H=-\mathbf{E}(t)\cdot\hat{\mathbf{d}}\,,\quad\hat{\mathbf{d}}=\frac{e}{2}\int d^{2}\mathbf{r}\mathbf{r}\hat{\Psi}^{\dagger}\tau_{z}\hat{\Psi}\,, \label{eq:deltaH}
\end{equation}
where $\mathbf{E}(t) = \mathbf{E}_0 \cos \omega t $ is the microwave-induced time-dependent electric field
 which is approximately uniform over the scale of the vortices~\footnote{In Eq.~(\ref{eq:deltaH}), we assume   a thin film superconductor that can be approximated by a 2D system. This model can also describe a 3D superconductor when the electric field $\mathbf{E}$ does not penetrate deep into its bulk. }. 
 The electric field couples to  the dipole operator  $\hat{\mathbf{d}}$ of the electronic states in the vortices. 
We have written it in terms of the  electron field operator in
Nambu spinor notation, $\hat{\Psi}=(\psi_{\uparrow},\psi_{\downarrow},\psi_{\downarrow}^{\dagger},-\psi_{\uparrow}^{\dagger})^{T}$; The Pauli matrix $\tau_z$ acts on the particle-hole indices. 
At low energies, we expand the field operators in terms of eigenstates as 
\begin{equation}
\hat{\Psi}(\mathbf{r})= \phi_{1}(\mathbf{r})\hat{\gamma}_{1}+\phi_{2}(\mathbf{r})\hat{\gamma}_{2}  +\Phi_{1}(\mathbf{r})\hat{\Gamma}_{1}+\Phi_{-1}(\mathbf{r})\hat{\Gamma}_{1}^{\dagger}+\dots \,, \label{eq:Psi}   \end{equation}
where $\hat{\gamma}_{1,2}$ are the Majorana operators for vortices 1 and 2, and $\hat{\Gamma}_{1}^{(\dagger)}$ is the annihilation (creation) operator for the lowest CdGM state. The corresponding wave functions multiply the operators in Eq.~(\ref{eq:Psi}).

At low frequencies much below the level spacing $\delta E$ of the vortex quasiparticle bound states, $\omega \ll \delta E /\hbar$, the microwave field does not excite the quasiparticle states of the vortices. 
We shall also assume that these quasiparticle states are not occupied, for example due to quasiparticle poisoning. 
Under these conditions, the vortex pair stays in its ground state manifold  consisting of the two states of unoccupied/occupied non-local MBS. 
With sufficiently weak microwave driving we can  use dispersive readout to measure the charge parity $\sigma_{z} =  i\hat{\gamma}_{1}\hat{\gamma}_{2} $~\cite{RevModPhys.93.025005,PRXQuantum.1.020313}. The dispersive Hamiltonian of the resonator-vortex pair system reads~\cite{PRXQuantum.1.020313}, 
\begin{equation}
H_\text{resonator} +   \delta H 
 = \hat{a}^\dagger \hat{a} (\hbar \omega + \sigma_{z} \hbar \chi) \,, \label{eq:MW+MZM}
\end{equation}
where $\hat{a},\hat{a}^\dagger$ are the  harmonic oscillator  annihilation and creation operators for the resonator. The MBS parity-dependent dispersive frequency shift is 
\begin{equation}
    \hbar \chi= \frac{g_1^2}{ \delta E}  \left[\frac{\delta E^2}{\delta E^2 - (\hbar \omega)^2} \right] \,, \label{eq:chi}
\end{equation} 
where we denote $g_1^2 = |\mathbf{E}_{0}\cdot\mathbf{d}_{1,- \mathcal{M}  }|^{2}-|\mathbf{E}_{0}\cdot\mathbf{d}_{1, \mathcal{M}}|^{2}$ and $\omega$ is the resonator bare frequency, $\mathbf{E}_0$ is the electric field amplitude, and $\delta E $  is the energy gap separating the MBS from the first excited CdGM mode. We ignore here the exponentially small energy splitting between the MBS, which would give subleading corrections to $\chi$; we will see that $\chi$ itself will be exponentially small in the vortex separation (due to the parity-sensitive transition dipole matrix elements $\mathbf{d}_{1,- \mathcal{M}  }$ and $\mathbf{d}_{1, \mathcal{M}  }$ being almost equal).  
We denote here $\mathbf{d}_{1,  \mathcal{M}  } = \langle 1 | \hat{ \mathbf{d}} | \mathcal{M} \rangle $ and $\mathbf{d}_{1, - \mathcal{M}  } = \langle \mathcal{M},1 | \hat{ \mathbf{d}} | 0 \rangle $ where the relevant state are the ground state $| 0 \rangle$, the single-particle excited states $ | \mathcal{M} \rangle = \hat{\Gamma}_{\mathcal{M}}^{\dagger} | 0 \rangle$ and $ | 1 \rangle = \hat{\Gamma}_{1}^{\dagger} | 0 \rangle$, and the two-particle excited state $ | \mathcal{M}, 1 \rangle = \hat{\Gamma}_{1}^{\dagger} \hat{\Gamma}_{\mathcal{M}}^{\dagger} | 0 \rangle$; we introduced the annihilation operator $\hat{\Gamma}_{\mathcal{M}}=(\hat{\gamma}_{1}+i\hat{\gamma}_{2})/2 $ for the non-local MBS. 

Evaluating the dipole transition matrix elements  $ \mathbf{d}_{1, \pm \mathcal{M}  }$ microscopically is somewhat involved since proper screening by the superconducting condensate  needs to be carefully accounted  for and is beyond the BCS mean-field theory~\cite{1996PhRvL..77..566B,2001PhRvL..86..312K,PhysRevB.91.045403,PhysRevB.97.125404,PhysRevX.8.031041}. 
Nevertheless, to estimate $\mathbf{d}_{1, \pm \mathcal{M}  }$ we can use Eq.~(\ref{eq:deltaH})  by replacing $\mathbf{r} \approx l \hat{\mathbf{z}}$ in it, with $l \approx a_B$  being the effective distance to the image charge in the superconductor and $\hat{\mathbf{z}}$ the surface  normal vector~\cite{1996PhRvL..77..566B}. Here $a_B$ denotes the Bohr radius. 
We evaluate the dimensionless matrix elements of the effective dipole ``charge'' $\mathbf{d}\cdot \hat{\mathbf{z}} / l$  by using a numerical simulation of the Majorana and CdGM states in a double vortex system depicted in  Fig.~\ref{fig:readout}b. The numerical simulations will be detailed in a future publication~\cite{MW_inprep}.

In Fig.~\ref{fig:readout}b we plot the parity-sensitive term $g_n^2$ that largely determines the dispersive shift $\chi$, Eq.~(\ref{eq:chi}). 
We find that even a relatively distant vortex pair can provide a  parity-dependent shift $g_n^2 \sim 10^{-2} (e l E_0)^2$. 
Since the relevant dipole moment is normal to the superconductor surface, we can couple to the dipole by using a microwave resonator above the surface,  producing a large perpendicular electric field. 
With a resonator zero-point voltage $V_0 \sim 100\, \mu \mathrm{V}$ at a $\sim  10 \mathrm{nm}$ distance from the vortices, we obtain $e l E_0 \approx 1 \mu \mathrm{eV} \cdot (l / \text{\AA}) \approx 2.4 \times 10^2 h \mathrm{MHz} \cdot (l / \text{\AA})$. (We estimate such high zero-point voltages can be achieved in high-inductance resonators~\cite{PhysRevApplied.5.044004}.) 
Taking a low-lying CdGM state with $\delta E \sim 10 \mu eV$, we obtain  $ \chi / 2\pi \sim   20 \,  \mathrm{MHz} \cdot (l / \text{\AA})^2$ where $l \sim a_B \gtrsim 1 \text{\AA}$ is the typical dipole size~\cite{1996PhRvL..77..566B}. 
We thus see that the MBS vortex parity measurement is well within standard circuit QED measurement  capabilities~\cite{RevModPhys.93.025005}. 
We note that the above estimate does not include the   resonant  enhancement, the second factor in Eq.~(\ref{eq:chi}), which may further substantially increase the frequency shift. 

Finally, we note that the dipole operator $\hat{\mathbf{d}}$ also has a non-zero diagonal matrix element $\mathbf{d}_{\mathcal{M}}$ in the Majorana state~\cite{PhysRevB.97.125404}, leading to a term $\mathbf{E}_0 \cdot\mathbf{d}_{\mathcal{M}} \sigma_z (\hat{a}+\hat{a}^\dagger)$ in Eq.~(\ref{eq:MW+MZM}). This term in principle allows one to perform longitudinal  readout of the MBS parity. However, making longitudinal readout practical may require parametric modulation of the coupling, in our case  $\mathbf{d}_{\mathcal{M}}$, which may be difficult~\cite{PhysRevB.99.235420,RevModPhys.93.025005}.   

\section{\label{sec:summ}Summary}

Measuring braiding statistics is the ultimate method to conclusively verify the existence of non-abelian excitations. We proposed a unified platform to initialize, braid and readout Majorana modes, avoiding abrupt topological-trivial interfaces at each stage. We derived general expressions for braiding speeds with spin currents, distance of closest approach between two Majorana modes and the resultant dispersive shift in cavity resonance frequency. We showed that our setup can be readily realized with existing options for TSC and MML materials.

\begin{acknowledgments}
We would like to thank Axel Hoffman and Mohammad Mushfiqur Rahman for helpful discussions. JIV thanks Dmitry Pikulin and Rafa\l{} Rechci\'{n}ski for helpful discussions on 3D TI simulations. JIV and LPR acknowledge support from the Office of the Under Secretary of Defense for Research and Engineering under award number FA9550-22-1-0354. YPC and PU acknowledges partial support of the work from US Department of Energy (DOE) Office of Science through the Quantum Science Center (QSC, a National Quantum Information Science Research Center) and NSF ECCS-1944635. STK acknowledges support from the Purdue research foundation.
\end{acknowledgments}

\appendix

\section{\label{app:A}Example Platform}

While our platform can be realized with a variety of existing combinations of MML and TSC, we choose Ir/Fe/Co/Ni and FeTe$_{0.55}$Se$_{0.45}$ for the respective roles\cite{petrovic2021skyrmion, wang2018evidence}. For the TSC, we used $\lambda = 500$ nm, $\xi=15$ nm and chose a thickness of $d_s = 50$ nm. This gives a Pearl length of $\Lambda = 5$ $\mu$m. We propose to use 10 layers of Ir/Fe/Co/Ni heterostructure with $0.5$ nm of Fe and $0.5$ nm of Co per layer giving an effective magnet thickness of $d_m = 10$ nm.

We performed micromagnetic simulations using mumax \cite{vansteenkiste2014design, mulkers2017effects} with parameters similar to those in Ref.~\cite{petrovic2021skyrmion}: $M_{0}=1450$ emu/cc, $A=13.9$ pJ/m, $K=2000$ kJ/m$^3$, $D=3$ mJ/m$^2$ and $B_{ext}=180$ mT, where $D$ is the Dzyaloshinskii–Moriya interaction (DMI) constant and $B_{ext}$ is the external magnetic field. We modified the values of uniaxial anisotropy constant and DMI constant from those in Ref.~\cite{petrovic2021skyrmion} to stabilize skyrmions of large size. We also used external magnetic field in the simulation to stabilize isolated skyrmions in the simulation. In experimental setup, this magnetic field can be applied via exchange interaction in magnetic multilayers. Thanks to versatile spintronics tools \cite{rana2020room, soumyanarayanan2017tunable, buttner2018theory, dupe2016engineering}, these modifications are quite feasible to realize. Using a grid size of $2\times2\times2$ nm$^3$, isolated skyrmions of size $r_{sk}=35$ nm were stabilized in the simulation.

To demonstrate skyrmion braiding, the MML was patterned into a Y-shaped track and a spin current corresponding to a charge current of $j = 2\times 10^{12}$ A/m$^2$ in HM layers was locally applied in each active leg of the MML track. To save computational resources, vacuum layers (representing heavy metal layers) have not been included in mumax simulations. The inclusion of vacuum layers does not affect the conclusions of this paper in any way. Here, we set spin-Hall angle, $\Theta _{SH}=0.15$, and the ratio of field-like SOT term to damping-like SOT term to $0.1$.

Analytical expression for nucleation condition of Eq.~(\ref{eqn:nuc}) was derived in Ref.~\cite{NeelInteraction} by assuming a linear ansatz for skyrmion profile where: $\alpha _{A} = 2\pi\int_{0}^{\infty} dx \,x [\theta^{\prime2}(x)+\sin^2\theta(x)/x^2]$, $\alpha _{K} = 4\pi\int_{0}^{\infty} dx \,x\sin^2\theta(x)$, and $\theta^\prime(x)=d\theta(x)/dx$, with $\theta(x) = \pi(1-x)$ for $x<1$ and $\theta(x) = 0$ otherwise. When the example material parameters are substituted in Eq.~(\ref{eqn:nuc}) with $r_{sk}=35$ nm, the left side of the equation exceeds the right side by about $400\%$, suggesting nucleation of a vortex-antivortex pair in the TSC for every skyrmion in the MML.

For quantitative analysis of Eqs.~(\ref{eqn:jminus})--(\ref{eqn:power}), we assume clean limit and neglect pinning forces on skyrmion and vortex. We include drag and Magnus forces on skyrmion, and drag force on vortex. We neglect Magnus force on the vortex as it is expected to be small compared to its drag force \cite{Checchin_2017}. We model drag force on the vortex using friction coefficient $\alpha_{vx}=\eta_{vx} d_s$, where $\eta_{vx}$ is the Bardeen-Stephen viscous drag coefficient given by $\eta_{vx} = \phi_0^2/(2 \pi \xi^2 \rho_n)$ \cite{PhysRevB.104.104501, PhysRev.140.A1197}. Here, $\rho_n$ is the normal state resistivity of the TSC. Following parameters were used for calculations: Gilbert damping parameter for the MML $\alpha=0.1$ and $\rho_n = 10^{-7}$ $\Omega$m giving a $\eta_{vx}=3\times10^{-8}$ Ns/m$^2$. We propose that each leg of the MML track in Fig.~\ref{fig:schematic} be $10 r_{sk}$ wide and $10 r_{sk}$ long. This implies that skyrmions have to cover a total distance of $50 r_{sk}$ for the entirety of the braiding process in Fig.~\ref{fig:braiding}. We used resistivity of Iridium $\rho_{hm}=5\times 10^{-7}$ $\Omega$m to calculate the Joule heat dissipated during the braiding process.

\bibliography{apssamp.bib}

\end{document}